# Critical transitions on route to chaos of natural convection on a heated horizontal circular surface

Yuhan Jiang[1,2], Yongling Zhao[2], Jan Carmeliet[2], Bingchuan Nie[1] and Feng Xu[1]
[1]School of Physical Science and Engineering, Beijing Jiaotong University, Beijing 100044, China
[2]Department of Mechanical and Process Engineering, ETH Zurich, Zurich, Switzerland

**Abstract** The transition route and bifurcations of the buoyant flow developing on a heated circular horizontal surface are elaborated using direct numerical simulations and direct stability analysis. A series of bifurcations, as a function of Rayleigh numbers ($Ra$) ranging from $10^1$ to $6\times10^7$, are found on the route to the chaos of the flow at $Pr=7$. When $Ra<1.0\times10^3$, the buoyant flow above the heated horizontal surface is dominated by conduction, because of which distinct thermal boundary layer and plume are not present. At $Ra=1.1\times10^6$, a Hopf bifurcation occurs, resulting in the flow transition from a steady state to a periodic puffing state. As $Ra$ increases further, the flow enters a periodic rotating state at $Ra=1.9\times10^6$, which is a unique state that was rarely discussed in the literature. These critical transitions, leaving from a steady state and subsequently entering a series of periodic states (puffing, rotating, flapping and doubling) and finally leading to chaos, are diagnosed using spectral analysis and two-dimensional Fourier Transform (2DFT). Moreover, direct stability analysis is conducted by introducing random numerical perturbations into the boundary condition of the surface heating. We find that when the state of a flow is in the vicinity of bifurcation points (e.g., $Ra=2.0\times10^6$), the flow is conditionally unstable to perturbations, and it can bifurcate from the rotating state to the flapping state in advance. However, for relatively stable flow states, such as at $Ra=1.5\times10^6$, the flow remains its periodic puffing state even though it is being perturbed.

**Key words:** Transition, buoyant flow, heated horizontal surface

## 1. Introduction

Natural convection is of longstanding interest in fundamental research of fluid mechanics (see e.g., Torrance et al. 1969; Worster 1986; Fan et al. 2021). In particular, the buoyancy-driven flows on heated horizontal surfaces, as a ubiquitous phenomenon, have been studied extensively in the past decades because of the underlying mechanisms in various nature and industrial systems, such as urban heat island (Fan et al. 2019), solar receiver (Samanes et al. 2015) and heat exchangers (Chen & Chou 2006). Specifically, plenty of literatures have concentrated on the dynamics and heat transfer of different regimes of the buoyancy-driven flows.

When a horizontal surface is heated, a thermal fluid layer appears and rises above it owing to the buoyancy that is gained through the heat transfer process, and the fluid layer is typically defined as a thermal boundary layer (TBL, Yu et al. 2007) and can be characterized by Grashof number ($Gr$) or Rayleigh number ($Ra$). Similarity solutions for the thermal boundary layer developed on the horizontal surface immersed in a fluid with Prandtl number ($Pr$) of 0.7 were firstly obtained using Boussinesq approximation by Stewartson (1958), in which an inclined angle $\alpha$ was used to identify the transverse pressure gradient induced by buoyancy. Gill et al. (1965) later revisited the boundary layer equations and indicated that the solutions are applicable when the heated surface faces upwards in a fluid. Several experiments have also been performed to investigate the flow pattern of the flows on heated surfaces of various geometries (e. g., square, rectangular, triangular and circular surfaces) (Husar & Sparrow 1968), the transition to large-eddy instability (Rotem & Claassen 1969), and the heat transfer (Pera & Gebhart 1973a). A summary of theoretical solutions and experimental correlations obtained in those studies has been made by Goldstein & Lau (1983), in which $Nu \sim Ra^{1/5}$ was observed in laminar flows where $Nu$ denotes Nusselt number.

As thermal boundary layers established from leading edges of a horizontal surface meet at its center, a starting plume accompanied by a large-eddy structure rises due to buoyancy (Rotem & Claassen 1969). Since the work of Batchelor (1954) and Morton et al. (1956) on the modelling of plumes, an increasing attention has been devoted to the study of plumes. Starting plumes, that rise above a horizontal surface in a two-dimensional model (Hier et al. 2004; Hattori et al. 2013b; Van Den Bremer & Hunt 2014), on a three-dimensional circular disk (Robinson & Liburdy 1987; Kitamura & Kimura 2008; Plourde et al. 2008; Lesshafft 2015; Kondrashov et al. 2016b; Sboev et al. 2018; Sboev & Kuchinskiy 2020), and from a line heat source (Noto 1989; Hattori et al. 2013a) have been studied using theoretical, experimental and numerical methods with a focus on the heat transfer and evolution of plume structures (e.g., cap and stem). Specifically, the influence of the geometry of heated surfaces on plumes (Kondrashov et al. 2016a, 2017) and other influencing factors such as ambient stratification (Torrance 1979; Lombardi et al. 2011; Marques & Lopez 2014; Mirajkar & Balasubramanian 2017) and Prandtl number (Worster 1986; Kaminski & Jaupart 2003) have been investigated extensively.

Scaling analysis is also adopted to obtain dominant scales of the dynamics and heat transfer of various transient flows in natural convection under different controlling parameters (Patterson & Imberger 1980; Lin et al. 2002; Armfield et al. 2007; Xu et al. 2009). The scaling laws for thermal boundary layers and plumes developed on a two-dimensional isothermally heated horizontal plate have been obtained in Jiang et al. (2019b), in which the scales for the steady state can be written as $\delta_T \sim W/Ra^{1/5}$ and $\delta_P \sim W/Ra^{4/15}$ respectively, where $\delta_T$ and $\delta_P$ denote the thickness of the thermal boundary layer and plume, and $W$ is the characteristic length of the horizontal plate. The heat transfer scaling relation reads $Nu \sim Ra^{1/5}$, which is consistent with the result presented in Goldstein & Lau (1983). The scales for the buoyant flow under isoflux and ramp heating conditions have also been revealed in Jiang et al. (2019a, 2021). The heat transfer relation in isoflux heating case is $Nu \sim Ra^{1/6}$, which is different from isothermal and ramp heating cases.

At small $Ra$, the buoyant flow above a horizontal surface is normally steady at equilibrium state after the starting plume rising upwards (Lopez & Marques 2013). With increasing Rayleigh number, the buoyant flow may exhibit different transitional flow structures. Vortical structures may form periodically along the plume axis, which triggers a periodic flow out of the steady flow (Hattori et al. 2013c). A rather intense flow may occur on the periphery of a heat source, instead of occurring at the surface center, in which symmetry of the plume is broken (Sboev et al. 2018). As the controlling parameters increase further, the vortices above the plate are amplified as they are convected towards downstream, which shed from the plate in the end and result in a fully turbulent flow (Kitamura et al. 2001).

Besides transient flows, the characteristics and structure of the flows at equilibrium state can lead to different flow states, which are quite interesting. The process of flow transition from steady to chaotic state is complicated, which can behave very differently in flows at different controlling parameters. With increasing controlling parameters such as *Ra*, the temporal and spatial complexity increases after a succession of bifurcations before the onset of turbulence (Drazin 2002). Ruelle & Takens (1971) and Newhouse et al. (1978) mapped a transition route using dynamic system theory, in which the steady flow became periodic, and it was followed by a quasi-periodic state exhibiting several different frequencies and finally it evolved into chaos. Libchaber & Maurer (1978) also suggested that a series of period-doubling bifurcations may exist when a flow evolves from a periodic state into a quasi-periodic state in a Rayleigh-Bénard convection model. The period of the flow is doubled as compared to that of the preceding state. The Ruelle-Taken-Newhouse route and the period-doubling transition are very common in many transitional flows, such as in the transition route of the flows in a cylinder cavity heated from below (Lopez & Marques 2013), on a top-open cavity heated at the bottom (Qiao et al. 2018), and adjacent to a heated section-triangular roof (Zhai et al. 2021).

The heat transfer of the transitional flows on the horizontal surface has been investigated as one of primary interest. Goldstein & Lau (1983) studied the heat transfer of the transitional flow on a heated horizontal plate using finite-difference analysis and experiments. Their results show that the power of the *Nu-Ra* correlation increases from 1/5 to 1/3 when the flow transits from laminar to turbulence, implying that the heat transfer is distinctly different in laminar and turbulent regimes. Their results are also consistent with other studies (see Clifton & Chapman 1969; Clausing & Berton 1989; Lewandowski et al. 2000; Wei et al. 2003).

Direct stability analysis has been used to study the stability of buoyant flows to understand the streamwise development of instabilities. The instability of travelling waves in the thermal boundary layer formed in a side-heated cavity was studied by adding perturbations to the start-up waves (Armfield & Janssen 1996). The stability features of steady-state flow were obtained and compared to start-up flow. The radiation-induced convective instability of fluid layer filled in a shallow wedge was studied by adding random and single-mode perturbations to the thermal boundary layer (Lei & Patterson 2003). The most unstable region of the fluid layer in the wedge and the most unstable mode of the convective instability were found. The critical time and Grashof number for the flow to become unstable were also obtained in good agreement with the scaling laws. Symmetric and asymmetric disturbances can be superimposed to a planar thermal plume on a finite heating source (Hattori et al. 2013b). The coupling of oscillations with the fundamental frequency and harmonics is observed in lapping flow and plume stem under different conditions. The study mainly focused on the development of oscillations in horizontal and vertical directions and also the relationship between the plume stem and lapping flow instabilities.

Although some states of the transitional flows on a heated horizontal surface are understood, the complete transition route from laminar to chaos along with the underlying bifurcations of the buoyancy-driven flows are rarely investigated. Kitamura & Kimura (2008) performed experiments with both water and air on the upward-facing circular disks in a wide range of *Ra*. The critical Rayleigh numbers for the beginning of the transition from laminar to turbulent flows were presented in two different working fluids. However, the study mainly focused on the heat transfer coefficients of the flows on the disks rather than the bifurcation routes of different states. Khrapunov & Chumakov (2020) performed experiments and numerical simulations of the flow on the horizontal surface with air as the working fluid, in which only results of one specific *Ra* have been presented and steady and puffing states have been found. For the study of swirling plume, Torrance (1979) observed a plume rotating about the centerline in a stratified cylindric enclosure in an experiment. It is well known that the swirling plume cannot be reproduced by two dimensional axisymmetric numerical simulations under the same conditions with experiments. Accordingly, Marques & Lopez (2014) investigated the bifurcations in a stratified ambient in a cylinder cavity similar with that in the experiment by Torrance (1979), and documented the presence of a swirling plume in three dimensional numerical simulations, which provides a new perspective to understand the naturally occurring swirling plumes. Unfortunately, the results in Marques & Lopez (2014) are restricted in a closed stratified ambient with linearly heating conditions, which is a quite small space, different from the stratification in atmosphere or ocean in nature.

In summary, the complete transition route, as a function of the controlling parameters of the buoyant flow on the isothermally heated horizontal circular surface, has not been investigated adequately in previous studies. The instability of transitional flows in different states is also rarely understood, which motivates our study. In this study, the complete transition route from steady to chaotic state of the buoyant flow on an isothermally heated horizontal surface is obtained with water (*Pr*=7) as the working fluid using three dimensional numerical simulations. Various bifurcations exhibiting different temporal and spatial characteristics are identified within a small range of *Ra*. Direct stability analysis is adopted subsequently to understand the stability of transitional states by introducing random numerical perturbations of different amplitudes. In the remainder of this paper, the physical problem and numerical methods are presented in § 2. A series of bifurcations of the buoyant flow on the horizontal surface, such as the steady state, periodic state, rotating state, period-doubling state and the influence of disturbance to the transitional flow, are discussed in § 3, followed by concluding remarks in § 4.

**2. Physical problem and numerical method**

2.1. Governing equations and controlling parameters

Transitional flow on a heated horizontal surface is simulated in the computational domain shown in figure 1. In this study, the three-dimensional continuity equation, momentum equations and energy equation with the Boussinesq assumption are used as governing equations. The non-dimensional governing equations can be written as below:

$$\frac{\partial u}{\partial x}+\frac{\partial v}{\partial y}+\frac{\partial w}{\partial z}=0, \quad (2.1)$$

$$\frac{\partial u}{\partial t}+u\frac{\partial u}{\partial x}+v\frac{\partial u}{\partial y}+w\frac{\partial u}{\partial z}=-\frac{\partial p}{\partial x}+\frac{Pr}{Ra^{1/2}}(\frac{\partial^2 u}{\partial x^2}+\frac{\partial^2 u}{\partial y^2}+\frac{\partial^2 u}{\partial z^2}), \quad (2.2)$$

$$\frac{\partial v}{\partial t}+u\frac{\partial v}{\partial x}+v\frac{\partial v}{\partial y}+w\frac{\partial v}{\partial z}=-\frac{\partial p}{\partial y}+\frac{Pr}{Ra^{1/2}}(\frac{\partial^2 v}{\partial x^2}+\frac{\partial^2 v}{\partial y^2}+\frac{\partial^2 v}{\partial z^2}), \quad (2.3)$$

$$\frac{\partial w}{\partial t}+u\frac{\partial w}{\partial x}+v\frac{\partial w}{\partial y}+w\frac{\partial w}{\partial z}=-\frac{\partial p}{\partial z}+\frac{Pr}{Ra^{1/2}}(\frac{\partial^2 w}{\partial x^2}+\frac{\partial^2 w}{\partial y^2}+\frac{\partial^2 w}{\partial z^2})+PrT, \quad (2.4)$$

$$\frac{\partial T}{\partial t}+u\frac{\partial T}{\partial x}+v\frac{\partial T}{\partial y}+w\frac{\partial T}{\partial z}=\frac{1}{Ra^{1/2}}(\frac{\partial^2 T}{\partial x^2}+\frac{\partial^2 T}{\partial y^2}+\frac{\partial^2 T}{\partial z^2}), \quad (2.5)$$

where $u$, $v$, and $w$ are velocity component in $x$, $y$, and $z$ direction in the Cartesian coordinate; $t$ is time; $p$ is pressure and $T$ is temperature. The governing equations are non-dimensionalized using the following scales: $x$, $y$, $z \sim D$; $u$, $v$, $w \sim \kappa Ra^{1/2}/D$; $\rho^{-1}\partial p/\partial x$, $\rho^{-1}\partial p/\partial y$, $\rho^{-1}\partial p/\partial z \sim \kappa^2 Ra/D^3$; $t \sim D^2/(\kappa Ra^{1/2})$ and $T-T_0 \sim \Delta T$. Here $D$ is the diameter of the horizontal surface; $\kappa$ is thermal diffusivity; $\rho$ is density; $\Delta T$ is temperature difference between the horizontal surface and the initial ambient fluid at $T_0$.

According to equations (2.2)–(2.5), the non-dimensional governing equations are dominated by two controlling parameters, which are Rayleigh number ($Ra$) and Prandtl number ($Pr$), defined respectively as shown below:

$$Ra = \frac{g\beta\Delta T D^3}{\nu\kappa}, Pr = \frac{\nu}{\kappa}, \quad (2.6)$$

where $g$ is gravitational acceleration; $\beta$ is coefficient of thermal expansion; $\nu$ is kinematic viscosity.

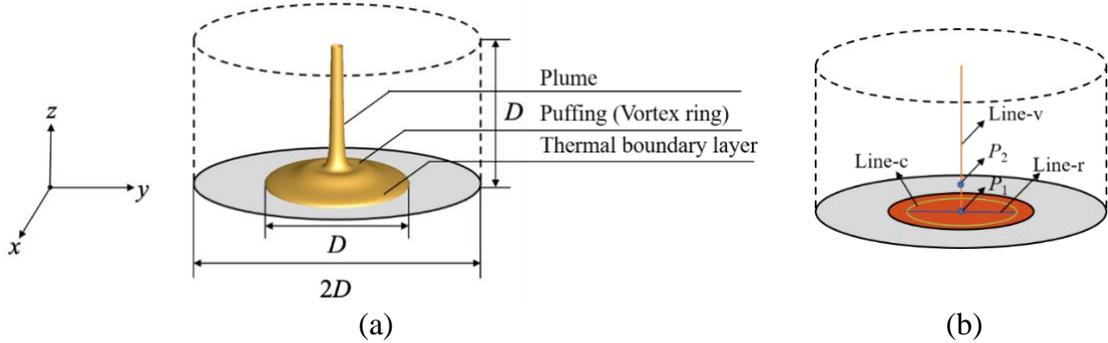

Figure 1 (a) Schematic of the computational domain with the temperature iso-surface ($T=0.1$) of a thermal plume in its periodic state, in which the center of the coordinate system aligns with the center of the heated surface. The dash lines denote open boundaries adopted in numerical simulations. (b) The probing points $P_1$ (0, 0, 0.01) and $P_2$ (0, 0, 0.2) and lines used in the following figures. The region highlighted in red denotes the heated surface; the gray part is the adiabatic wall. The line-v denotes a vertical line leaving the center of the heated surface; line-r and line-c denote radial and circumferential lines, respectively.

## 2.2. Geometry, boundary conditions and grid

In this study, the working fluid is water ($Pr=7$), which is treated as incompressible Newtonian fluid. As shown in figure 1(a), the flow develops on a heated circular rigid surface of diameter $D$ that is enclosed by a cylindrical open boundary with a diameter of $2D$ and a height of $D$.

The following initial and boundary conditions are adopted to investigate the transition and instability of the thermal boundary layer and plume developing on an isothermally rigid horizontal surface. Here, the bottom boundary is modelled as a non-slip and rigid wall. An isothermal condition of $T=1$ is imposed on the central circular surface for $t>0$, and an adiabatic condition of $\partial T/\partial y=1$ on the annular extension. The top and side boundaries of the computational domain are set as open boundaries using pressure outlet conditions, where $T$ is zero and Neumann conditions with zero normal derivative are applied to the velocity components. At the initial time, the temperature and velocity of all the boundaries and the working fluid are zero.

All boundary and initial conditions can be expressed as:

$$\begin{cases} u = v = w = 0,\ T = 1,\ \text{for}\ x^2 + y^2 \leq \dfrac{1}{2},\ z = 0\ \text{and}\ t > 0 \\ u = v = w = \dfrac{\partial T}{\partial z} = 0,\ \text{for}\ \dfrac{1}{2} < x^2 + y^2 \leq 1,\ z = 0\ \text{and}\ t > 0 \\ \dfrac{\partial u}{\partial y} = \dfrac{\partial v}{\partial y} = \dfrac{\partial w}{\partial y} = T = 0,\ \text{for}\ x^2 + y^2 = 1,\ 0 < z < 1\ \text{and}\ t > 0 \\ \dfrac{\partial u}{\partial z} = \dfrac{\partial v}{\partial z} = \dfrac{\partial w}{\partial z} = T = 0,\ \text{for}\ z = 1\ \text{and}\ t > 0 \\ u = v = w = T = 0,\ \text{for}\ t = 0 \end{cases} \quad (2.7)$$

## 2.3. Numerical simulation methods

The governing equations are solved using the finite volume method with SIMPLE scheme. The advection term is discretized by QUICK scheme; the diffusion term is discretized by second-order central difference scheme; the transient term is discretized by second-order backward implicit time-marching scheme. This numerical approach has been used successfully in other studies (Qiao et al. 2018; Jiang et al. 2021).

For discretization of the domain, an O-Type multi-grid system is established at the Cartesian coordinates in which finer grids are applied close to boundaries, and coarse grids in other regions. Such a non-uniform grid system can be used to satisfactorily resolve the regions where larger velocity and temperature gradients are expected to occur.

A series of tests are carried out to find the optimal time step, grid and domain size to ensure computing accuracy and the use of affordable computing resources. Three different grid systems with 0.5 million, 1 million and 2 million cells along with three non-dimensional time steps of 0.0005, 0.001, 0.002 are tested for the case of $Ra=6\times10^7$. In $x$- and $y$-direction, the grid system is constructed with $\Delta x$ and $\Delta y=0.007$ with an expansion factor of 1.04 from the center to the boundary edge. In $z$-direction, the grid is constructed with $\Delta z=0.007$ with an expansion factor of 1.06 from the bottom to the upper boundary. The temperature and velocity in the cases with different grid systems and time steps are monitored at point $P_1$ (0, 0, 0.01), as illustrated in figure 1(b). The average temperature and velocity as well as the relative variations with the reference solution of test 2 are listed in table 1. Based on the results in table 1, the variation between test 3 and the reference test 2 is 0.8%, and the one between test 2 and test 1 is 2.3%. Furthermore, the variation between test 2 and test 4 and test 5 is small. Therefore, the case of test 2 with the grid system of 1 million cells and time step of 0.001 are adopted as the best option given both computing accuracy and cost.

| Case No. | Number of cells (million) | Time step | z-velocity | Temperature |
| --- | --- | --- | --- | --- |
| test 1 | 0.5 | 0.001 | 0.01771 (2.3%) | 1.927 (3.1%) |
| test 2 | 1 | 0.001 | 0.01813 | 1.989 |
| test 3 | 2 | 0.001 | 0.01828 (0.8%) | 2.003 (0.7%) |
| test 4 | 1 | 0.0005 | 0.01833 (1.1%) | 1.956 (1.7%) |
| test 5 | 1 | 0.002 | 0.01846 (1.8%) | 2.019 (1.5%) |

Table 1 Average $z$-velocity and temperature at point $P_1$ (0, 0, 0.01) for $Ra=6\times10^7$ using different number of cells and time steps.

In order to make sure that the open boundaries do not influence the buoyant flow, tests of different sizes of the computational domain are conducted. That is, the size of the domain is diameter of $2D$ and height of $D$ in test 2, diameter of $2D$ and height of $2D$ in test 6, and diameter of $4D$ and height of $2D$ in test 7. As shown in table 2, all variations of the temperature and velocity between reference test 2 and other tests is less than 1%. Since the flow in the thermal boundary layer and the plume stem is near the surface center, the domain of $2D\times D$ is considered to be large enough to capture the characteristics and structure of the transitional flow. Accordingly, under consideration of accuracy and cost, the domain size of $2D\times D$ is adopted in the study.

| Case No. | Domain size (Diameter×height) | z-velocity | Temperature |
| --- | --- | --- | --- |
| test 2 | $2D\times D$ | 0.01813 | 1.989 |
| test 6 | $2D\times 2D$ | 0.01808 (0.3%) | 1.979 (0.5%) |
| test 7 | $4D\times 2D$ | 0.01829 (0.9%) | 2.010 (1%) |

Table 2 Average $z$-velocity and temperature at point $P_1$ (0, 0, 0.01) for $Ra=6\times10^7$ using different domain sizes.

For a further validation, figure 2 shows numerical results calculated using the present numerical simulation method and experimental data from Khrapunov et al. (2017). That is, the profile of the local Nusselt number along the disk radius is calculated and presented at $Ra=3.78\times10^6$ (or $Gr=5.40\times10^6$). Clearly, the numerical results agree well with experimental data from Khrapunov et al. (2017), suggesting that the present numerical method is guaranteed to describe the buoyant flow on the surface.

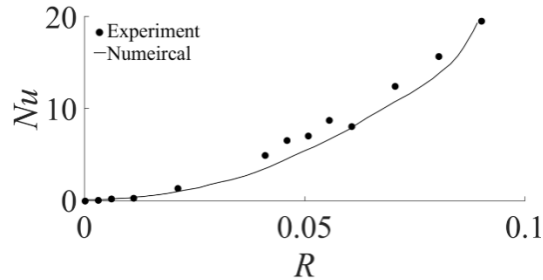

Figure 2 Validation of the local Nusselt number from present numerical method with experimental results from Khrapunov et al. (2017).

## 3. Results and discussion

To unravel the transition route of the buoyant flow on the horizontal surface as complete as possible, hundreds of cases for $Ra=10^1-6\times10^7$ were calculated to distinguish various bifurcations, flow structures and corresponding bifurcation points (critical Rayleigh number).

With increasing $Ra$, the buoyant flow on the horizontal surface is observed to go through a series of bifurcations, starting with the conduction dominated flow, followed by a steady flow, a periodic flow with different states such as puffing, rotating and flapping, then a period-doubling flow and finally transition to chaos. Since a rotating state found in the transition route was rarely discussed in previous studies, we performed direct stability analysis to further understand the rotating state. Random perturbations were imposed on the heated surface to understand the response of the buoyant flow in different transitional states.

3.1. The transition route to chaos
3.1.1. Primary flow

Starting off with $Ra=10^1$, heat transfer is dominated by conduction from the heated surface to the fluids in a very weak flow. As shown in figure 3, there is no distinct thermal boundary layer or plume, but a steady axisymmetric dome structure. According to the group theory (Crawford & Knobloch 1991), if a system remains unchanged under arbitrary rotation around the central axis and with reflections over any vertical plane through the central axis, the symmetry group of solutions to the governing equations under such boundary conditions is termed the group of $O(2)$. With increasing $Ra$ from $10^1$ to $10^3$, a noticeable convection starts to establish, which leads to the shrinking of the dome structure, where the high temperature region contracts and concentrates towards the center of the dome.

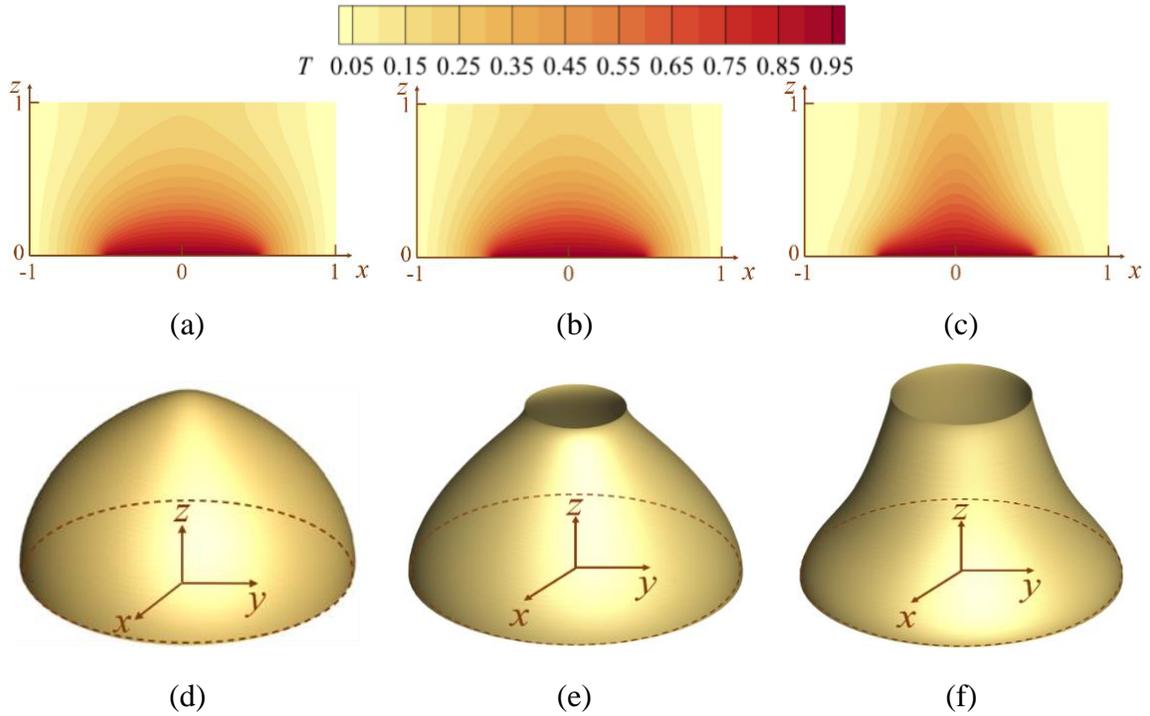

Figure 3 The *x-z* plane temperature contour plot of conduction-dominated flow at *t*=50 for (a) $Ra=10^1$, (b) $Ra=10^2$ and (c) $Ra=10^3$; and corresponding temperature iso-surface at *T*=0.1 and *t*=50 for (d) $Ra=10^1$, (e) $Ra=10^2$ and (f) $Ra=10^3$.

For relatively larger *Ra*, the convection effect becomes much more pronounced, which takes place over the dominance of the conduction effect. The thermal boundary layer and starting plume can be readily distinguished in figure 4. The radial temperature gradient and baroclinic effect drive the fluid radially inward and rise upward due to the buoyancy effect, finally forming a distinct plume. The temperature of the plume decreases with increasing height because of the dissipation of heat from the plume to the ambient fluid in the process of plume rising. With increasing *Ra*, the thickness of the thermal boundary layer and plume reduces, respectively. This is because the convection effect becomes stronger, leading to a larger temperature gradient. The heat transfer in this case is more intense, which is compatible with the scaling laws obtained for a heated horizontal surface model in Jiang et al. (2019b). As shown in figure 4, the flow is still steady and axisymmetric.

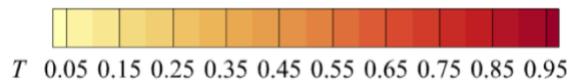

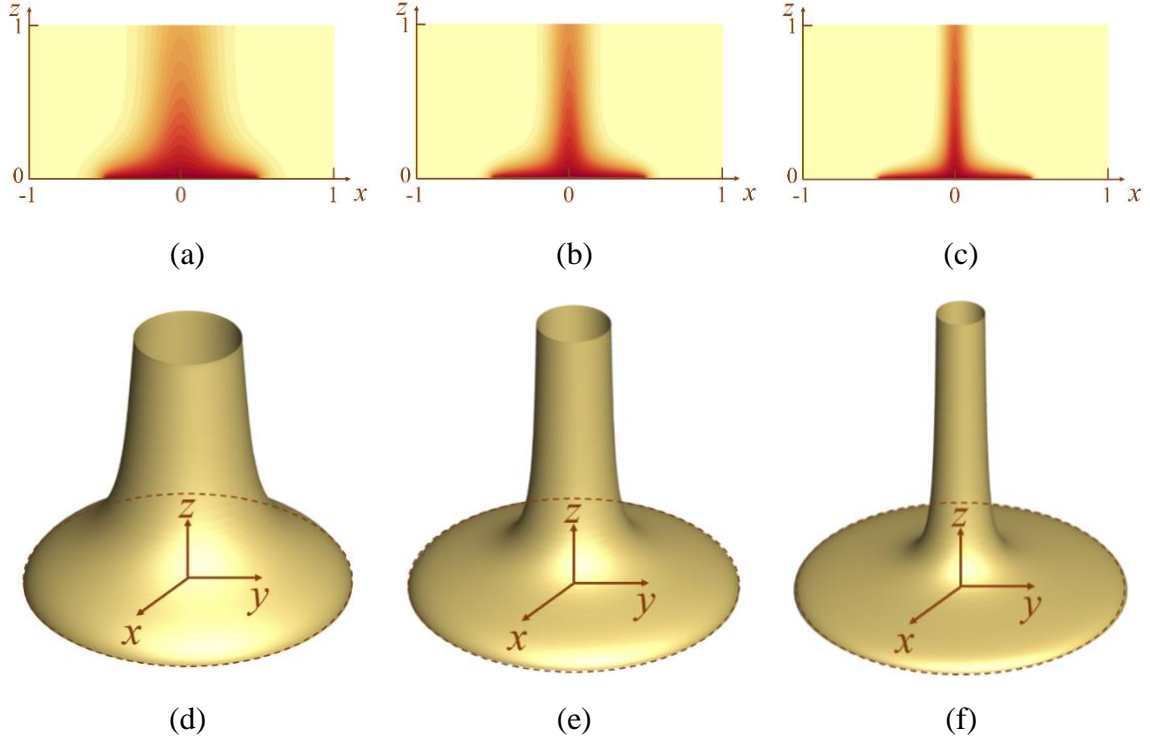

Figure 4 The *x-z* plane temperature contour plot of convection-dominated flow at $t=50$ for (a) $Ra=10^4$, (b) $Ra=10^5$, (c) $Ra=10^6$; and temperature iso-surface at $T=0.1$, $t=50$ for (d) $Ra=10^4$, (e) $Ra=10^5$, (f) $Ra=10^6$.

3.1.2. Hopf bifurcation: periodic puffing flow

With further increasing *Ra*, a Hopf bifurcation occurs, triggering the buoyant flow to transit from steady to periodic state. The flow structures characterized by temperature contours of a complete cycle in *x-y* and *x-z* planes at different times are shown in figure 5 and the times are indicated using dots in figure 6(b), from which the evolution of the thermal boundary layer and plume can be readily observed. As *Ra* increases, the radial flow in the thermal boundary layer driven by baroclinic effect speeds up and the fluid carried away by the rising plume is less than the fluid accumulated due to the radial flow, which leads to the formation of a vortex ring. Then, the vortex ring sheds from the thermal boundary layer, and it is entrained by the rising plume. After the departure of the vortex ring, the flow from the edge of the plate fills in so that the process repeats over time, resulting in a periodic flow that is called 'puffing' state, as illustrated in List (1982).

As shown in figure 5(a), the flow structure in puffing state is an axisymmetric plume. A puffing forms in the thermal boundary layer on the outer side of the thermal plume in *x-z* plane, which is produced by the strong buoyant flow from the edge of the heated circular surface in figures 5(b). The puffing is finally convected to the rising plume, as shown in figure 5(c). In fact, the puffing evolves into a vortex ring in the thermal boundary layer around the thermal plume as shown in figure 5(f). As the puffing moving towards the thermal plume, the vortex ring also shrinks in *x-y* plane as depicted in figures 5(f)−(g). According to the temperature contour plot in *x-z* plane, the puffing forms periodically on the edge of the thermal boundary layer, which can be illustrated from the multiple vortex rings in *x-y* plane. The vortex ring forms at the edge of the circular surface and moves towards the center. Finally, it merges into the plume due to the entrainment of the plume and flows upward with the plume. At this state, the group of *O(2)* still exists, which means it is in an axisymmetric state. The temperature contour and isosurface can be referred to movie 1 and 2.

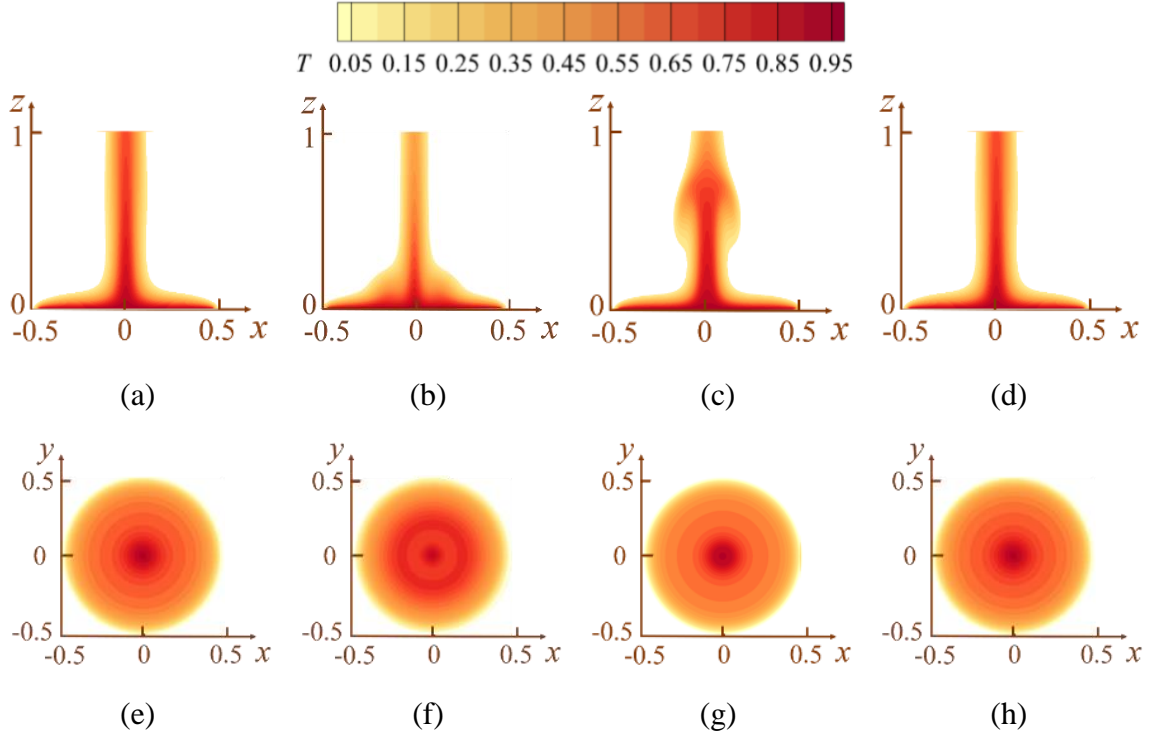

Figure 5 The *x-z* plane temperature contour plot for $Ra=1.1\times10^6$ at equilibrium state for one period at (a) *t*+5.3, (b) *t*+5.7, (c) *t*+6.4, (d) *t*+7.3; The *x-y* plane temperature contour plot at height of $0.01D$ for $Ra=1.1\times10^6$ at equilibrium state for one period at (e) *t*+5.3, (f) *t*+5.7, (g) *t*+6.4, (h) *t*+7.3.

Figures 6(a) and (b) plot the *z*-direction velocity (*w*) time series at point $P_1$ for the flow at $Ra=1.0\times10^6$ and $Ra=1.1\times10^6$, respectively. While the velocity still remains constant at $Ra=1.0\times10^6$, it becomes periodic at $Ra=1.1\times10^6$, implying that the solution loses its stability and bifurcates into a periodic solution by means of Hopf bifurcation. Additionally, the power spectrum density is plotted, confirming that the frequency is in the periodic state, as shown in figure 6(c). The buoyant flow is periodic and exhibits a fundamental frequency (non-dimensionalized by $(\kappa Ra^{1/2})/D^2$) $f_f=0.412$, where some harmonic modes are also seen. As the period of the velocity time series only exhibits one peak, and thus the power spectrum density of the fundamental frequency and harmonics decay linearly.

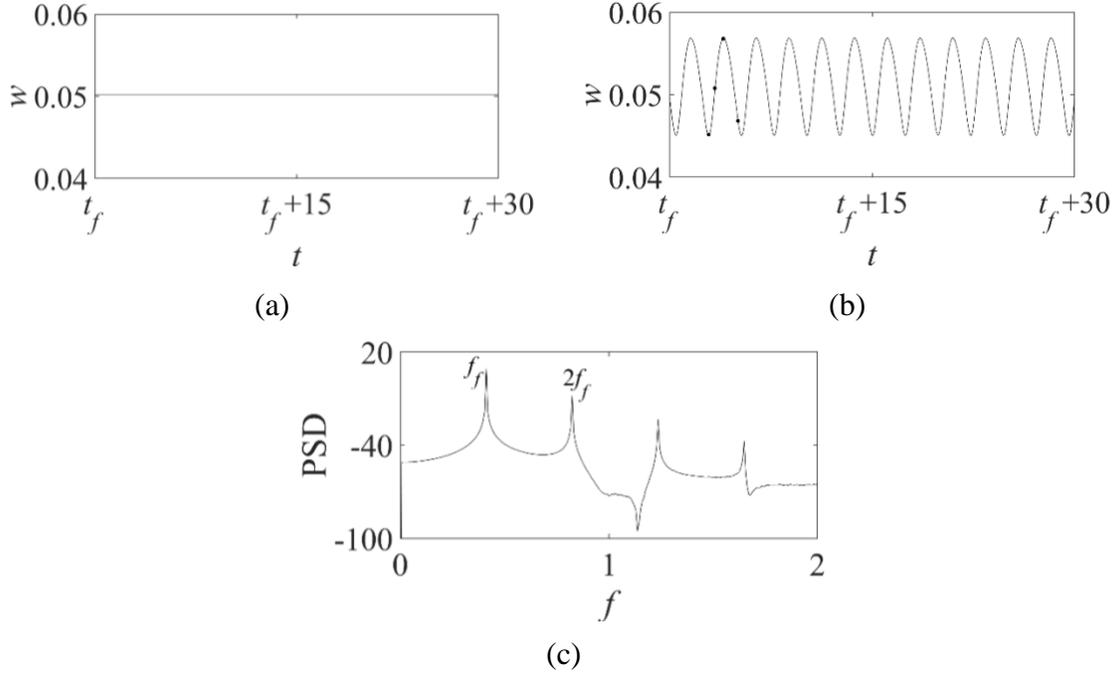

Figure 6 (a) Time series of $z$-velocity $w$ at $P_1$ for $Ra=1.0\times10^6$, (b) Time series of $z$-velocity $w$ at $P_1$ for $Ra=1.1\times10^6$, the black dots are the time instants that used in figure 5, (c) Power spectrum density of the periodic flow for $Ra=1.1\times10^6$ where $f_f=0.412$.

3.1.3. Symmetry-breaking Hopf bifurcation: periodic rotating flow

In this study, when $Ra$ increases to $1.9\times10^6$, the periodic puffing state becomes unstable and a different periodic solution presented as a rotating state is observed. As shown in figure 7, the plume begins to rotate in anti-clockwise order, as a consequence of which the flow becomes asymmetric. In this rotating state, the puffing does not appear simultaneously in the thermal boundary layer as they do in the puffing state. Instead, it appears on one side and rotates around the $z$-axis. The plume in the center has quite a large heat flux going upwards at this $Ra$, which entrains the puffing around it. That is, the asymmetry of the puffing leads to this type of the rotating plume. The temperature contour and isosurface can be referred to movie 3 and 4.

Clearly, when an equilibrium state undergoes a symmetry-breaking bifurcation, new fluid flow states appear with less symmetry and more sophisticated dynamics. According to the study of Crawford & Knobloch (1991), breaking $O(2)$ symmetry can lead to either standing or rotating waves when the bifurcating eigenvalue is complex. Furthermore, because of the reflection symmetry, the rotating state can be in clockwise or anticlockwise direction; both solutions bifurcate simultaneously, one of which can be observed, dependent on initial conditions. Most rotating flows are generated through symmetry breaking, as already observed in convection in a closed cavity (Murphy & Lopez 1984; Navarro & Herrero 2011).

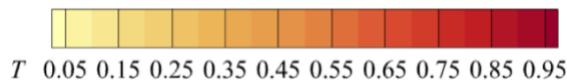

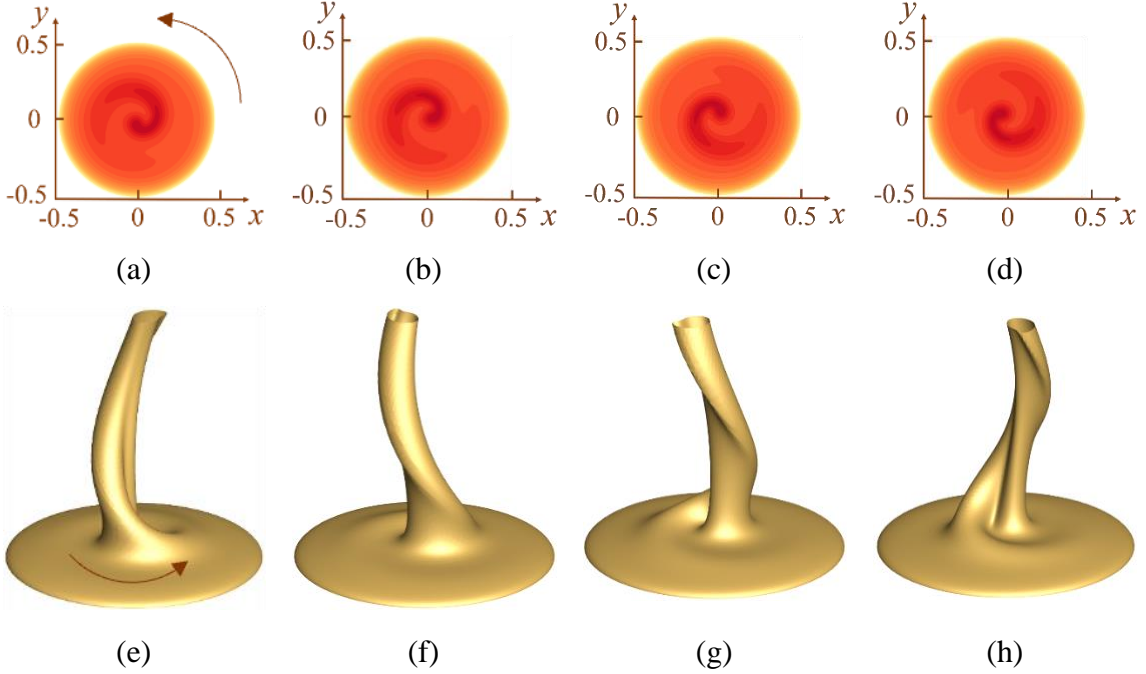

Figure 7 The *x-y* plane temperature contour plot at height of 0.01*D* for $Ra=2\times10^6$ at equilibrium state for one period at (a) $t+4.6$, (b) $t+5.2$, (c) $t+5.8$, (d) $t+6.4$; The temperature iso-surface for $T=0.1$ and for $Ra=2\times10^6$ at equilibrium state for one period (e) $t+4.6$, (f) $t+5.2$, (g) $t+5.8$, (h) $t+6.4$. The arrow denotes that the flow rotates in anti-clockwise direction.

The circumferential velocity can be used to distinguish the axisymmetric flow. When the circumferential velocity at one point inside the plume but away from the vertical axis equals to zero, the flow has no velocity in circumferential direction and thus only moves directly upwards without rotation. When the plume begins to rotate in a specific direction, the circumferential velocity will be non-zero. In figure 8, the circumferential velocity is near zero when *Ra* is smaller than $1.7\times10^6$, but it increases slightly to $1.7\times10^{-4}$ at $Ra=1.8\times10^6$, suggesting that the solution tends to become unstable and potentially approaches the critical bifurcation point. After the circumferential velocity suddenly rises to 0.018 at $Ra=1.9\times10^6$, the flow becomes asymmetric and this symmetry-breaking process is responsible for the propagation of rotation. It is worth noting that there are two orders of magnitude differences of the circumferential velocity between $Ra=1.8\times10^6$ and $1.9\times10^6$. The circumferential velocity at $Ra=1.8\times10^6$ is quite small, which is probably because of the oscillations at the bifurcation point. So, we may conclude that the rotation state begins at $Ra=1.9\times10^6$. After that, the circumferential velocity decreases again but will not be zero, that's mainly because the flapping flow in 3.1.4 still has circumferential velocity but not as strong as rotating flow.

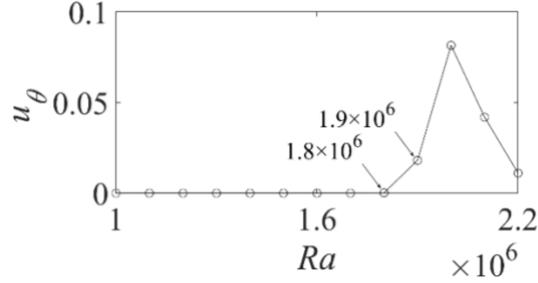

Figure 8 Velocity in circumferential direction for different Rayleigh numbers.

As displayed in figure 9(a), the $z$-velocity time series of the rotating state is quite complex and different from that of the puffing state, in which the amplitude of the $z$-velocity oscillations is small compared with the puffing state. This is mainly because the plume is rotating away from the center as shown in figure 7 but the $z$-velocity is monitored in the center, and in turn the amplitude of the $z$-velocity oscillations is smaller. Figure 9(b) shows the power spectrum density for $Ra=2.0\times10^6$. Clearly, the flow in the rotating state is still periodic with discernable harmonic modes. The power spectrum density of the rotating mode has multiple spike frequencies in which the power of harmonic frequencies is even larger than that of the fundamental frequency. This is mainly because that the $z$-velocity has multiple oscillations in each rotating period, as shown by the velocity plot in figure 9(a); in particular, four discernable oscillations in one rotating period may result in the power spectrum density with one fundamental frequency and three distinct harmonic frequencies, as displayed in figure 9(b).

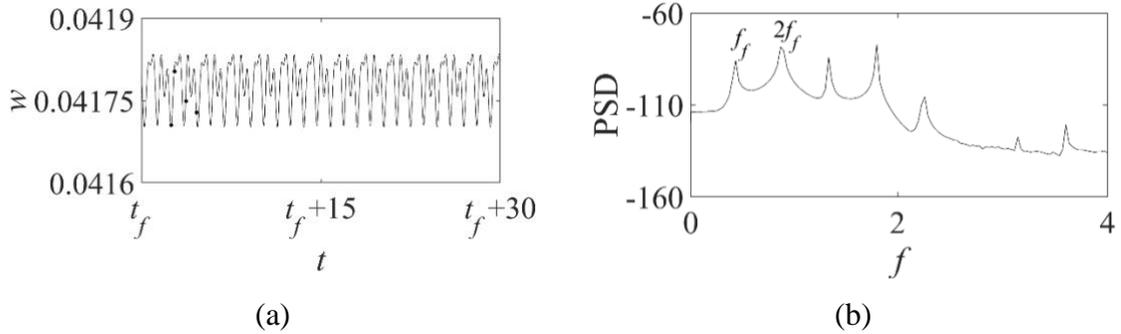

(a)          (b)

Figure 9 (a) Time series of $z$-velocity $w$ at $P_1$ for $Ra=2.0\times10^6$, the black dots are the time instants that used in figure 7, (b) power spectrum density of the periodic flow for $Ra=2\times10^6$ where $f_f=0.434$.

3.1.4. Reflection-symmetric Hopf bifurcation: periodic flapping flow

With increasing $Ra$, a Hopf bifurcation occurs between $Ra=2.1\times10^6$ and $Ra=2.2\times10^6$, in which the periodic rotating state breaks but a periodic flapping state with reflection symmetry occurs. The flow structure after the Hopf bifurcation is also different from the structures presented in figure 5. Figures 10(a)–(h) show that the puffing forms alternately at the 'two sides' (outer region) of the plume in the thermal boundary layer; that is, the vortex ring is not symmetric, different from those in the periodic puffing state in figure 5.

The tilted puffing makes the plume deviating away from the z-axis at the center, leading to a 'flapping' plume. The flapping of the plume around its axis can also explain why there is a velocity reduction followed by a rise in one period in the velocity time series. When the plume flaps away from the center, the velocity in the center decreases; when the puffing moves towards the plume, the velocity increases again.

It is worth noting that, in the flapping state, the reflection symmetry is still preserved, while the rotational symmetry has been broken. The buoyant flow is symmetrical with the vertical plane through the central axis, as illustrated by the dash line in figures 10(e)–(h). That is, this kind of asymmetrical vortex ring makes the plume flapping in a certain direction as shown in the top view of the temperature contours in x-y plane. The temperature contour and isosurface can be referred to movie 5 and 6.

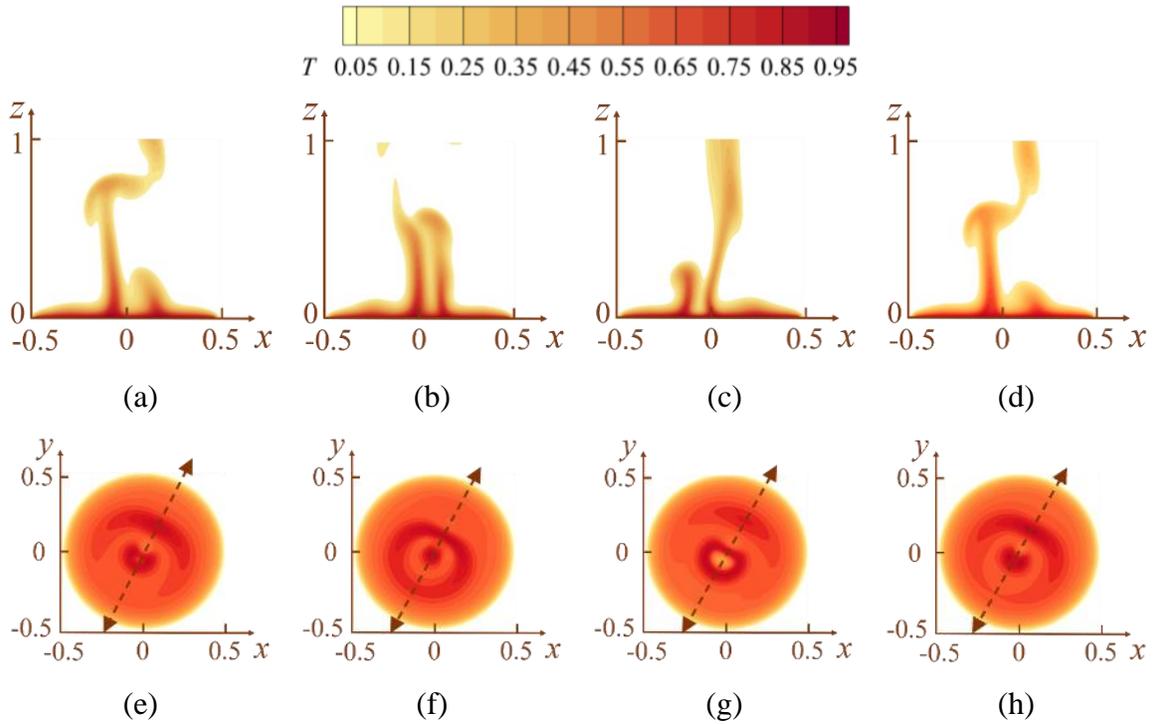

Figure 10 The x-z plane temperature contour plot for $Ra=2.5\times10^6$ at equilibrium state for one period at (a) $t+3.8$, (b) $t+4.4$, (c) $t+5.0$, (d) $t+5.6$; The x-y plane temperature contour plot at height of $0.01D$ for $Ra=2.5\times10^6$ at equilibrium state for one period at (e) $t+3.8$, (f) $t+4.4$, (g) $t+5.0$, (h) $t+5.6$. The arrow denotes that the flow flaps in this direction.

Figure 11 plots the z-velocity time series and also the power spectrum density in the flapping state. The flow is still periodic with distinct fundamental and double frequencies. However, the fundamental frequency becomes larger ($f_f=0.465$) than that of the previous states ($f_f=0.412$). That is, the fundamental frequency slightly grows with increasing Rayleigh number. The spectrum of other harmonics decays as the increasing of frequency, which is also the convergence characteristic of the periodic flow.

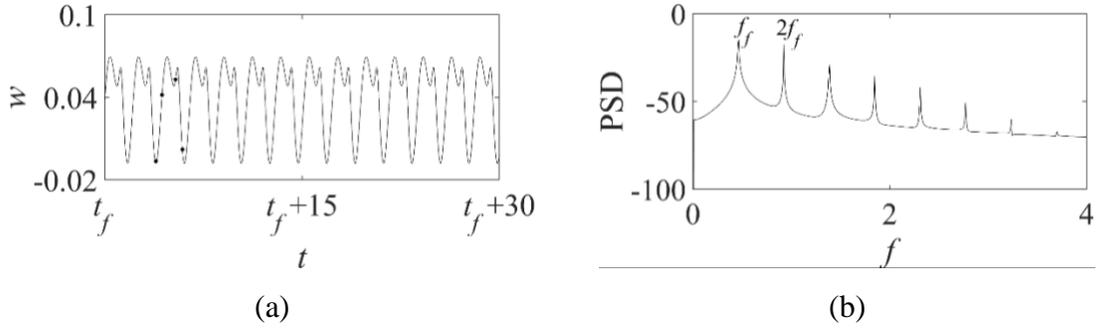

(a)                                      (b)

Figure 11 (a) Time series of $z$-velocity $w$ at $P_1$ for $Ra=2.5\times10^6$, the black dots are the time instants that used in figure 10, (b) power spectrum density of the periodic flow for $Ra=2.5\times10^6$ where $f_f=0.465$.

3.1.5 Period-doubling bifurcation

Figure 12 shows the temperature at $Ra=6.5\times10^6$. Clearly, the second ring (VR2) interacts with the first ring (VR1) before the first one vanishes. That is, there are two small periods in one complete cycle, which is the period-doubling bifurcation. The flow structures are depicted for the period-doubling state in figure 12. The flow remains a flapping state, but sways in a different direction compared to that in the periodic state in e.g., figure 10. However, the difference between the two swaying direction is quite small and hard to be distinguished in these figures, but is carefully compared and verified. The swaying direction of the plume mainly depends on the initial conditions. However, the flapping plume does not sway on the whole heated plate like it does in the periodic flapping state, but only sways in a small region as shown in figures 12(a)–(d). That is mainly because with the Rayleigh number increasing, the characteristic length of the heated surface that we observed in the flow field also becomes bigger. Thus, the flapping on the whole heated plate shrinks into a small region. Additionally, two puffings form successively and merge into one puffing near the center, which is different from the periodic flapping state. According to the temperature contour in figures 12(e)–(h), two vortex rings (VR1 and VR2) exist simultaneously and are then entrained by the plume and flow upwards, which may explain why the period doubles in this state.

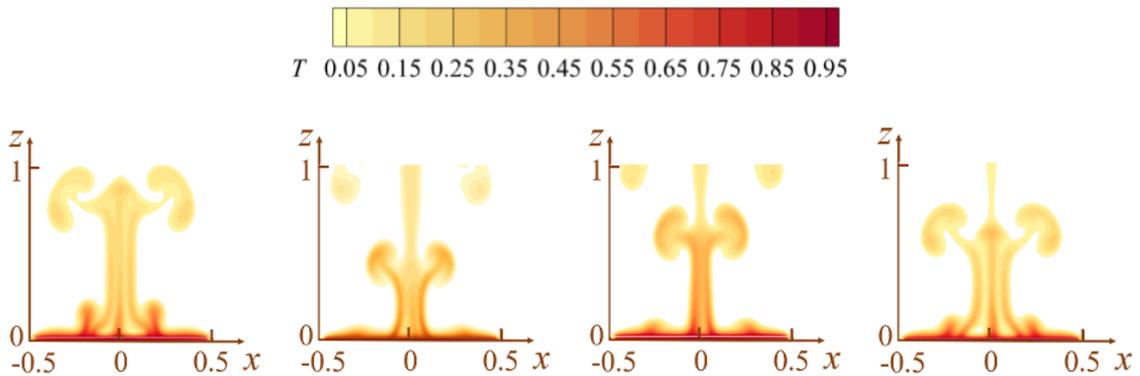

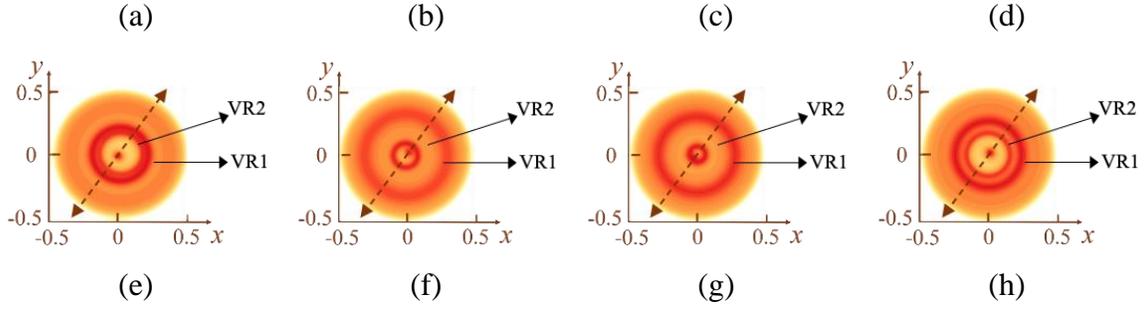

(e)                (f)              (g)              (h)

Figure 12 The *x-z* plane temperature contour plot for $Ra=6.5\times10^6$ at equilibrium state for one period at (a) $t+2.9$, (b) $t+5.2$, (c) $t+7.5$, (d) $t+9.8$; The *x-y* plane temperature contour plot at height of $0.01D$ for $Ra=6.5\times10^6$ at equilibrium state for one period (e) $t+2.9$, (f) $t+5.2$, (g) $t+7.5$, (h) $t+9.8$. The vortex rings are pointed out in the plot as VR1 and VR2

Figure 13 shows the *z*-velocity time series and the corresponding power spectrum density at $Ra=6.5\times10^6$. The variance between two neighbor waves is very minor compared to the amplitude of the velocity oscillations, as revealed in the zoom in figure 13(a). The minor variance implies that the period doubling bifurcation occurs, as illustrated by the half of the fundamental frequency in figure 13(b). According to figures 12(e)-(f), the plume always sways in a certain direction. As a result, the variance of the horizontal velocity is larger than that of the vertical velocity in some case.

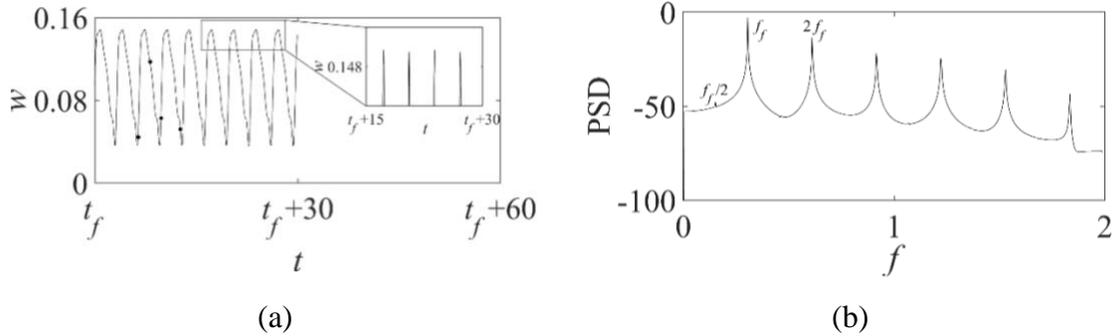

(a)                                 (b)

Figure 13 (a) Time series of *z*-velocity $w$ at $P_1$ for $Ra=6.5\times10^6$, the black dots are the time instants that used in figure 12, (b) power spectrum density of the periodic flow for $Ra=6.5\times10^6$ where $f_f=0.305$.

Figure 14 shows the temperature contours at $Ra=3\times10^7$. The buoyant flow above the surface at $Ra=3\times10^7$ becomes axisymmetric again; that is, the plume in the center does not sway any more. With further increasing Rayleigh number to $3\times10^7$, the flow is still in a period-doubling state. However, the flow structure differs from that at $Ra=6.5\times10^6$. The puffing forms simultaneously on the outer sides and then merges, before it is convected upwards by the plume, as shown in figures 14(a)−(d). That is, the vortex rings form at the edge of the heated surface and move inwards to the center axis, which remains symmetric, as shown in figures 14(e)–(h). The flow structure will be distinguished and analyzed thoroughly in § 3.3.

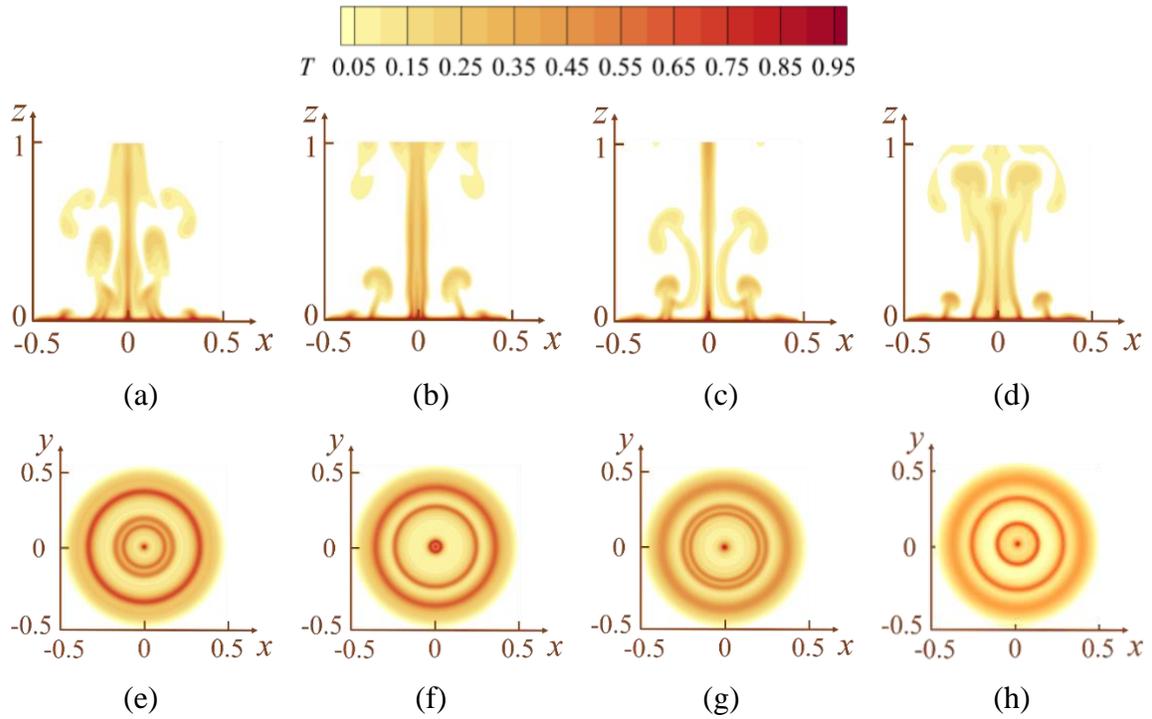

Figure 14 The *x-z* plane temperature contour plot for $Ra=3\times10^7$ at equilibrium state for one period at (a) $t+3.0$, (b) $t+5.4$, (c) $t+7.8$, (d) $t+10.2$; The *x-y* plane temperature contour plot at height of $0.01D$ for $Ra=3\times10^7$ at equilibrium state for one period at (e) $t+3.0$, (f) $t+5.4$, (g) $t+7.8$, (h) $t+10.2$.

3.1.6. Transition to chaos

The buoyant flow above the horizontal surface becomes more complex and finally evolves into a chaotic state. Figure 15 displays the temperature contours of the chaotic state at $Ra=5.02\times10^7$ in *x-z* plane. The plume in the center is quite chaotic and presents flow structures of various length scales. However, the puffing formed at the edge of the plate still exhibits similarities to some extent, because the local *Ra* at the edge of the circular plate is smaller than that in the center. Thus, the flow may still present some periodic characteristics.

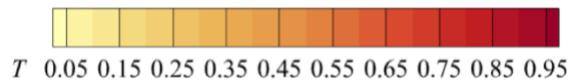

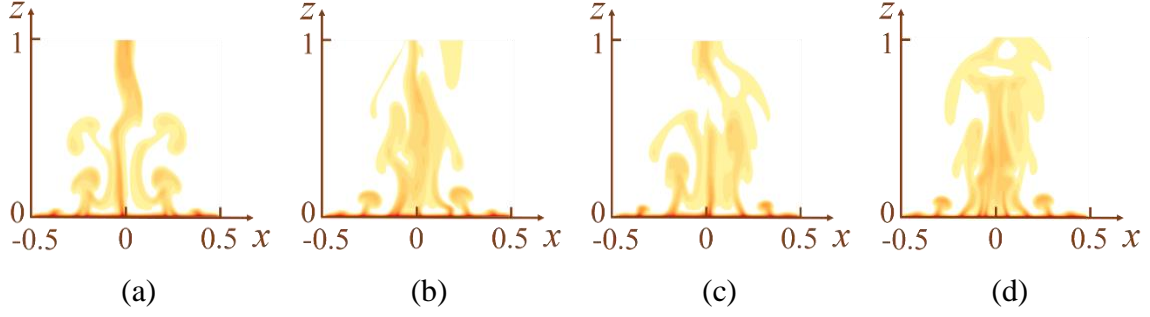

Figure 15 The *x-z* plane temperature contour plot for $Ra=5.02\times10^7$ at equilibrium state at (a) $t+2.0$, (b) $t+3.5$, (c) $t+5.0$, (d) $t+7.5$.

Figures 16(a) and (b) plot the *z*-direction velocity time series and power spectrum density at $Ra=5.02\times10^7$. Clearly, the *z*-velocity becomes more complex without presenting any clear period in figure 16(a) and there is also no clear fundamental frequency in the power spectrum density in figure 16(b), suggesting that the flow is entering chaotic. However, the flow still has some major frequencies that can be distinguished, which implies that the flow is not in a fully-developed chaotic state. This state is only slightly beyond the period-doubling state, and thus still has some characteristics in common with the period-doubling state.

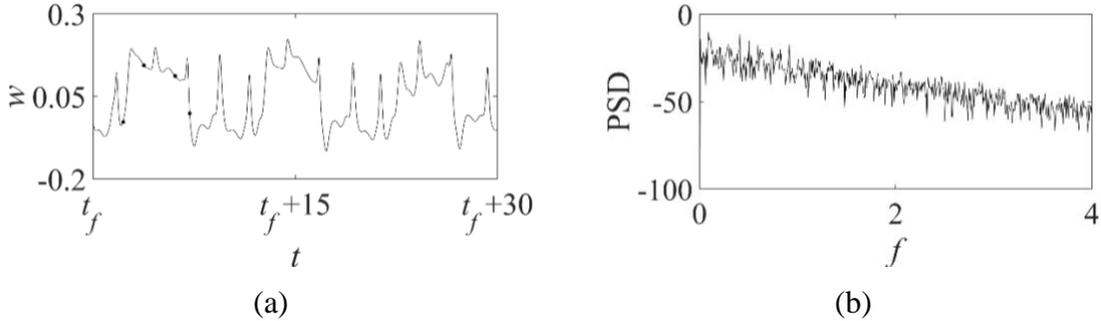

Figure 16 (a) Time series of *z*-velocity *w* at $P_1$ for $Ra=5.02\times10^7$, where the black dots are the time instants that used in figure 15, and (b) power spectrum density of the periodic flow for $Ra=5.02\times10^7$.

3.1.7. The whole route to chaos

For better understanding the route to a chaotic state, the trajectories in the state space *u-w-T* are demonstrated in figure 17 for the typical flow states on the circular surface. The trajectory finally approaches a fixed point at $Ra=1.0\times10^6$ at the steady state in figure 17(a) and a limit cycle at $Ra=1.1\times10^6$ at the periodic puffing state in figure 17(b). For the periodic rotating and flapping state at $Ra=2.0\times10^6$ and $Ra=2.5\times10^6$, the trajectory is also a limit circle, but to some extent twisted compared with the puffing case (figures 17(c) and (d)). A $T^2$ torus appears at $Ra=3.0\times10^7$ in which the flow is in the period-doubling state in figure 17(e). Finally, the trajectory becomes a complex attractor at $Ra=5.02\times10^7$ in which the flow is chaotic, as shown in figure 17(f).

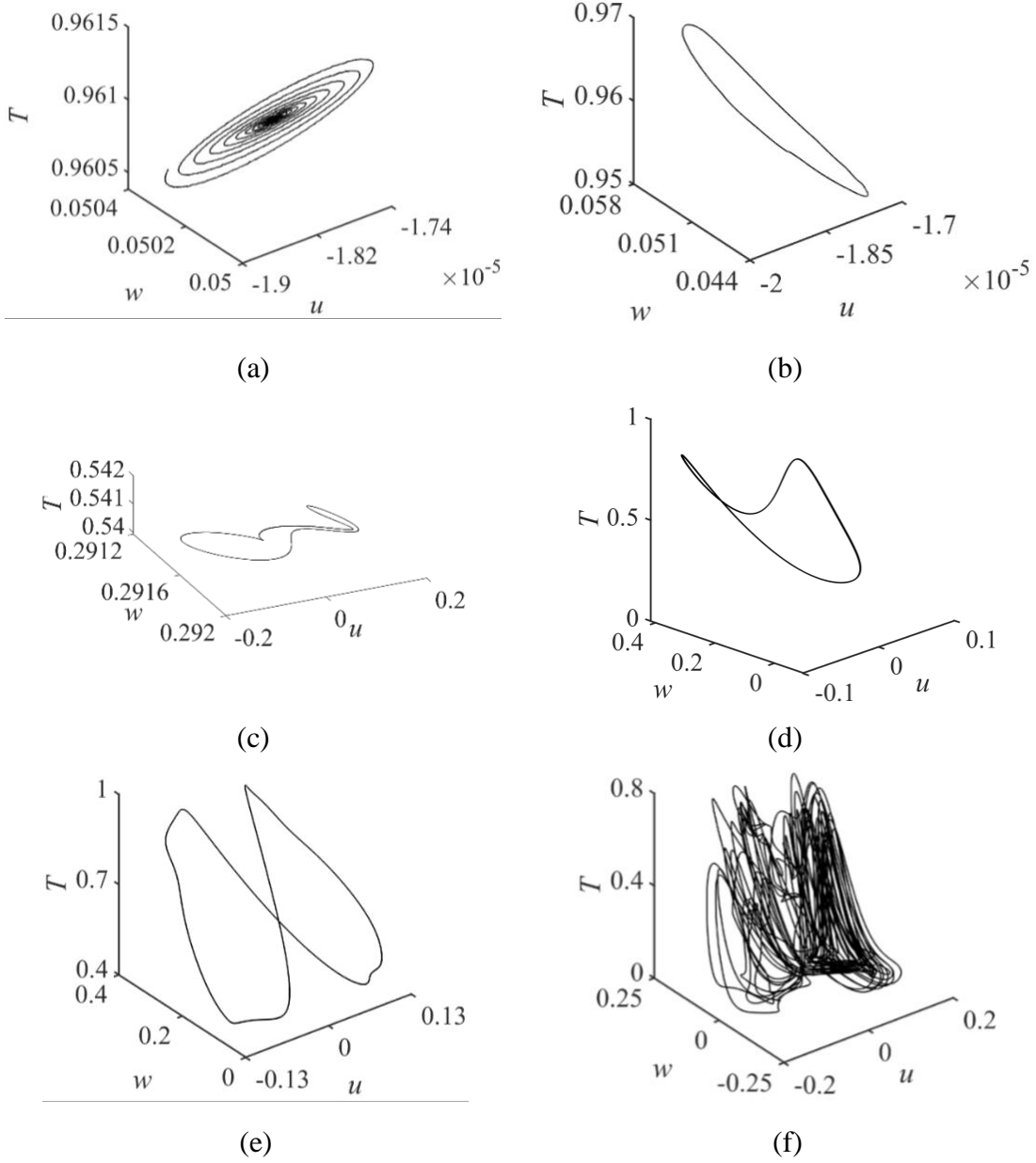

Figure 17 The attractors of *x*-velocity *u*, *z*-velocity *w* and temperature *T* at (a) $Ra=1.0\times10^6$, (b) $Ra=1.1\times10^6$, (c) $Ra=2.0\times10^6$, (d) $Ra=2.5\times10^6$, (e) $Ra=3.0\times10^7$, (f) $Ra=5.02\times10^7$.

To identify the chaotic state, the maximum Lyapunov exponent ($\lambda_L$) of the attractors in figure 17 is calculated, defined as (Odavić et al. 2015)

$$\lambda_L = \frac{1}{t_n - t_0} \sum_{i=1}^{n} \ln \frac{d_1 t_i}{d_0 t_i}. \qquad (2.8)$$

In this equation, $d_0$ is the initial distance of two points selected in the orbit of the attractors (may also see figure 17). For the next time $t_1=t_0+\Delta t$ (e.g., $\Delta t=100$ time steps), the two points arrive at new positions, and thus the distance between two points becomes $d_1(t_1)$. Further, we can find new points with a distance $d_0$, and then we may start the next calculation and iterate several times ($n>50$). As shown in figure 18, $\lambda_L$ becomes larger than zero for $Ra \geq 5.02\times10^7$, beyond which the flow enters the chaotic state (also see Odavić et al. 2015 for chaos described by the maximum Lyapunov exponent).

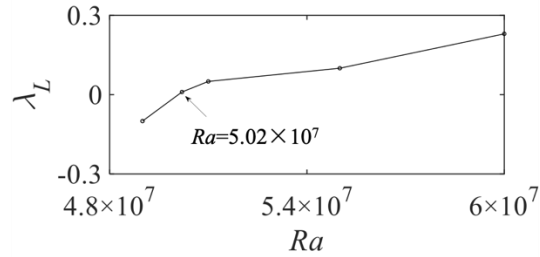

Figure 18 The maximum Lyapunov exponent $\lambda_L$ for different Rayleigh numbers.

For purpose of illustrating the 3D flow explicitly, the time series of temperature profile are presented in vertical axial, radial and circumferential directions, as shown in figure 19. As illustrated in figure 1(b), the vertical line-v originates from the original point and ends at the center of the top boundary. The radial line-r goes across the original point and its length is $4/5D$. The circumferential line-c diameter is also $4/5D$. Line-r and line-c are both $1/30D$ away from the bottom boundary. Note that at this height, the temperature profile can describe the structures of both thermal boundary layer and thermal plume, which is more complete. Especially, in the circumferential direction, $\theta$ refers to the non-dimensional angle of the circle from 0 to 360 degrees. For better understanding, we trim and stretch the circle into a straight line and depict the temperature profile along it. In different directions, the variation of complex flow structures with increasing Rayleigh number can be distinguished clearly.

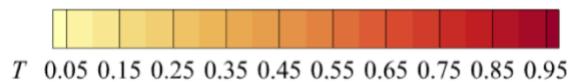

$Ra=1.0\times10^6$

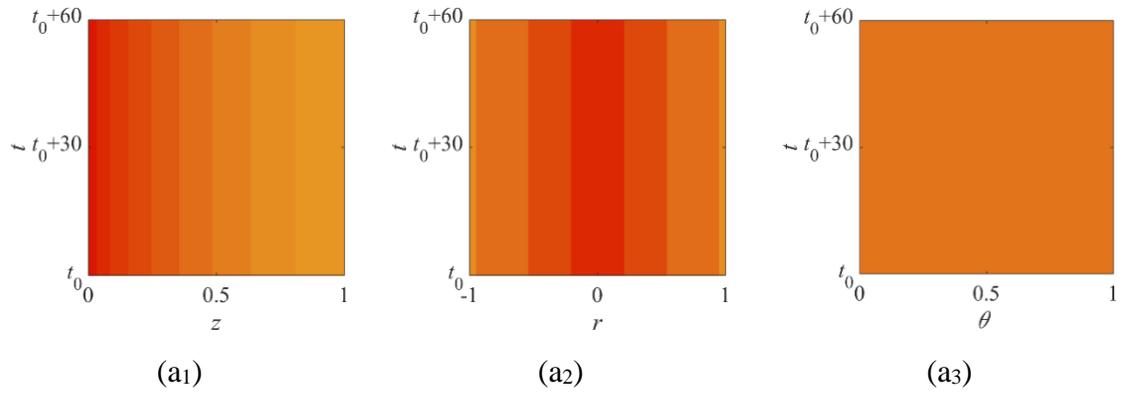

(a₁)　　　　　　　　(a₂)　　　　　　　　(a₃)

$Ra=1.1\times10^6$

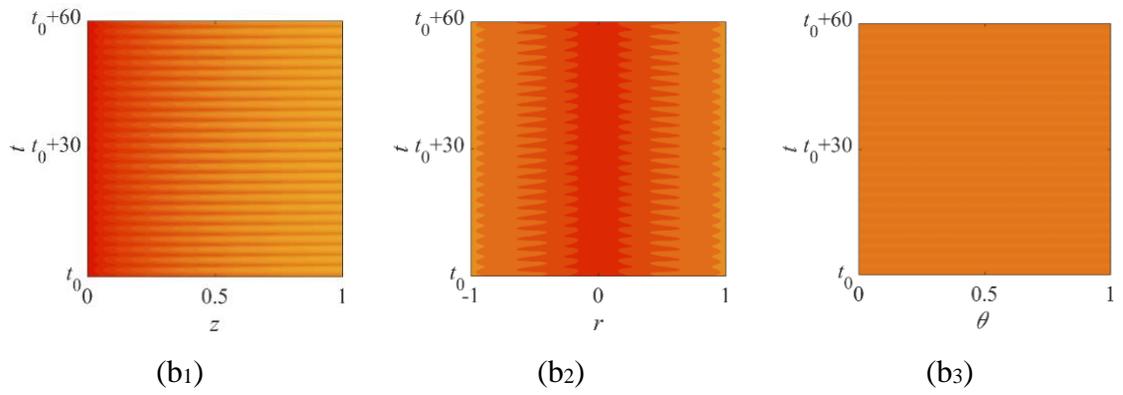

(b₁)　　　　　　　　(b₂)　　　　　　　　(b₃)

$Ra=1.9\times10^6$

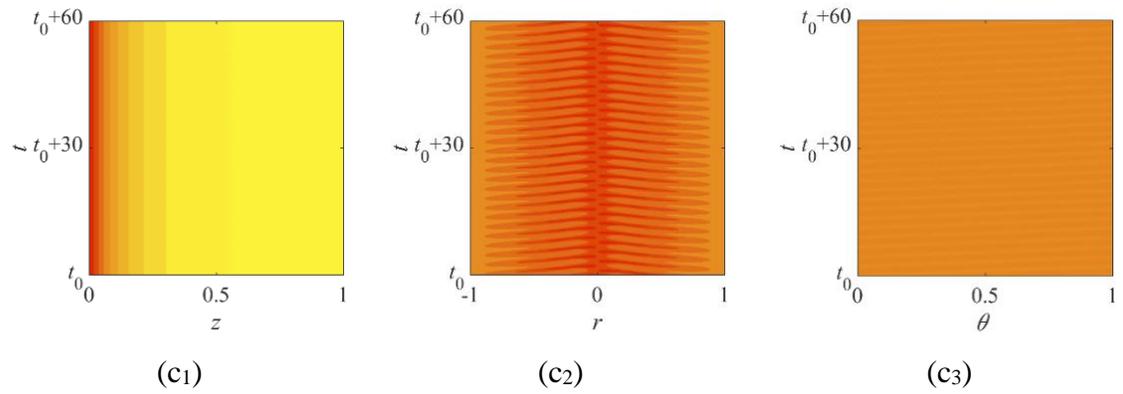

(c₁)　　　　　　　　(c₂)　　　　　　　　(c₃)

$Ra=2.3\times10^6$

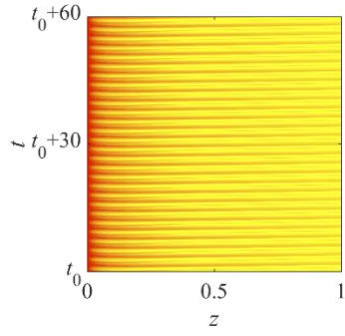
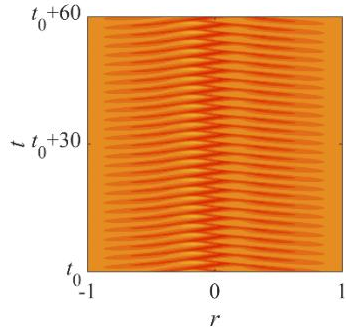
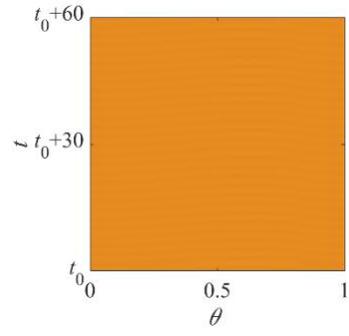

(d₁) (d₂) (d₃)

$Ra=6.5\times10^6$

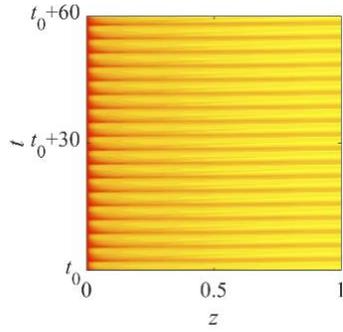
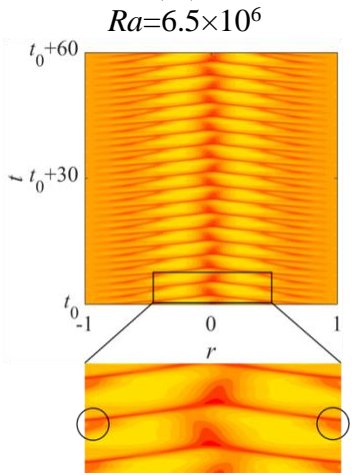
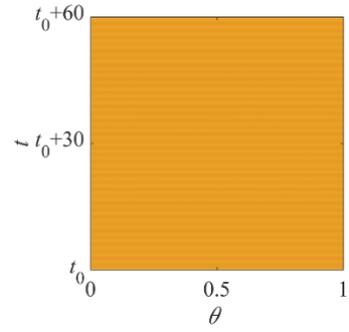

(e₁) (e₂) (e₃)

$Ra=3.0\times10^7$

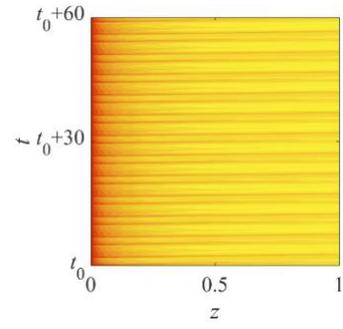
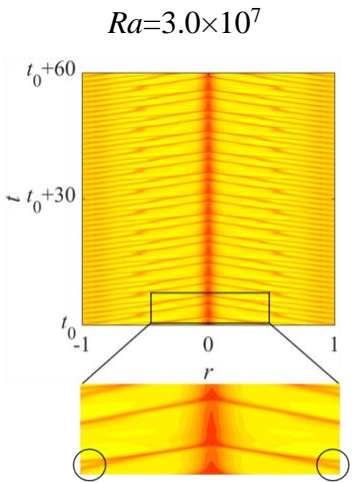
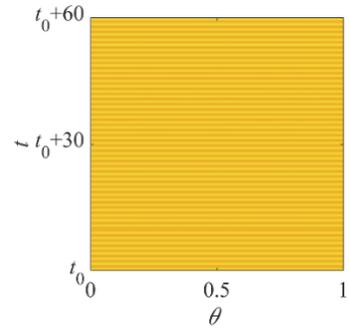

(f₁) (f₂) (f₃)

$Ra=5.02\times10^7$

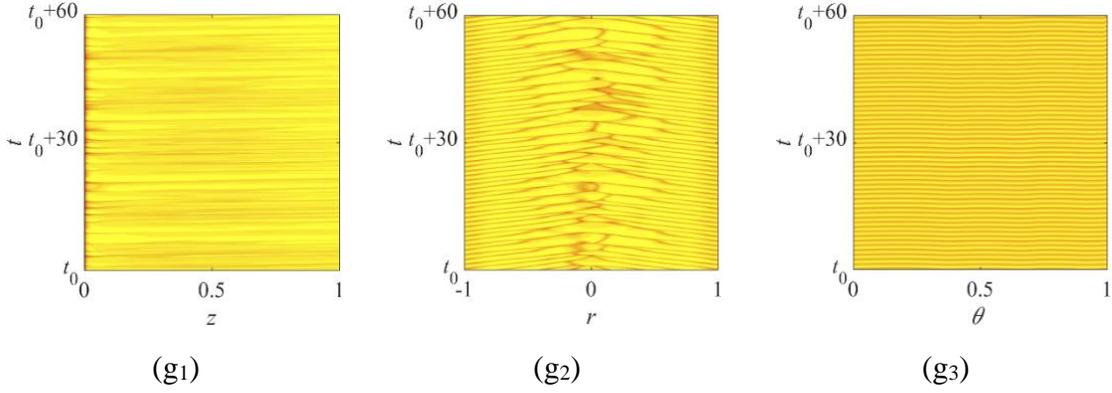

(g₁)   (g₂)   (g₃)

Figure 19 Temperature time series in vertical, radial and circumferential direction at (a$_1$) (a$_2$) (a$_3$) $Ra=1.0\times10^6$, (b$_1$) (b$_2$) (b$_3$) $Ra=1.1\times10^6$, (c$_1$) (c$_2$) (c$_3$) $Ra=1.9\times10^6$, (d$_1$) (d$_2$) (d$_3$) $Ra=2.3\times10^6$, (e$_1$) (e$_2$) (e$_3$) $Ra=6.5\times10^6$, (f$_1$) (f$_2$) (f$_3$) $Ra=3.0\times10^7$, (g$_1$) (g$_2$) (g$_3$) $Ra=5.02\times10^7$.

When $Ra=1.0\times10^6$, the flow is in a steady state. In the vertical direction, the temperature declines with increasing height. The heat is transferred from the heated surface into the surrounding fluid by convection and conduction. In figures 19(a$_1$) and (a$_2$), the temperature decreases from the center to the edge of the surface in radial direction because the horizontal flow in the thermal boundary layer is continuously heated by the surface, and from the surface to the top in vertical axial direction because of the continuous heat loss of the rising plume to the ambient fluid. However, the temperature is still uniform in the circumferential direction in figure 19(a$_3$), because the flow is axisymmetric. Additionally, the temperatures remain constant with time as the flow is in the steady state.

At $Ra=1.1\times10^6$, the flow is not steady any more. As shown in figure 19(b$_1$), the temperature becomes periodic in the vertical direction. The magnitude of the temperature fluctuates, which peaks when the puffings merge into the plume and drops when the plume rises up. Periodic stripes appear with time in figure 19(b$_2$), implying that the flow is in a periodic state. According to the radial temperature profile, the stripes (puffings) appear simultaneously on both sides of the radial line. In circumferential direction the temperature is constant on the circle in the same time, proving it is also a symmetric state.

With increasing $Ra$ to $1.9\times10^6$, the flow bifurcates into a rotating state by means of a symmetry-breaking Hopf bifurcation. As shown in figure 19(c$_1$), the temperature decreases sharply in vertical axial direction and the puffing stripes can be hardly distinguished. That is mainly because in the rotating state, the plume rotates around the center, rather than stationarily appearing at the center location. Thus, the temperature gradient is quite large in the center. Given the temperature profile in the other two directions, the puffing appears alternately on different sides of the radial line, which means the flow is no longer symmetric, as shown in figure 19(c$_2$). Further, as depicted in figure 19(c$_3$), the temperature at the end of this period is the same as that at the beginning of the next period, which suggests the plume rotates around the circle.

At $Ra=2.3\times10^6$, the flow structure changes again after going through a Hopf bifurcation. The stripes can be distinguished clearly in the vertical direction, as shown in figure 19($d_1$). The temperature difference between different stripes is also larger than that in the puffing state due to the flapping of the plume. When the plume sways to the edge of the heated surface, the temperature at the center line decreases significantly. In radial direction, the puffing appears alternately on different sides of the radial line, and the plume in the center sways to the left and right to interact and merge with the puffing on different sides, which is referred to as a flapping state. In figure 19($d_3$), the temperature in circumferential direction is not spatially homogenous at the same time. The stripes turn into a 'wave' shape, and the wave evolves with the change of the time, suggesting that the flow is periodic but asymmetric.

At $Ra=6.5\times10^6$, the period-doubling bifurcation occurs. As shown in figure 19($e_1$), the period is apparently longer than that of the periodic flow shown in figure 19($b_1$). Two stripes appear in one period, which can be described as a period-doubling state. A different type of flow structure appears in radial direction. First, a puffing appears at the edge of the heated surface and moves to the plume. Before it arrives at the center, another puffing also appears and moves just like the previous one. Two puffings encounter and merge into one puffing, moving towards the plume and finally being entrained by the plume. As shown in figure 19($e_2$), two stripes encounter in the process of flowing towards the center and become one stripe. The plume in the center sways to its left and right, which makes this state an asymmetric flow. This special flow structure is the combination of the puffing and flapping state. In circumferential direction, the temperature profile is more like an axisymmetric puffing mode, as shown in figure 19($e_3$). That is mainly because the circle chosen in the circumferential direction is close to the edge of the heated horizontal surface, where the flow structures captured on this circle is only the puffing flow away from the center.

The flow is still a period-doubling flow at $Ra=3.0\times10^7$. However, the flow structure becomes quite different. As shown in figure 19($f_1$), the two stripes in vertical direction are clearer compared to those at $Ra=6.5\times10^6$, which means that the plume does not sway away from the center but always puffs in the center. Thus, the two stripes are very similar in one period. In radial direction, the puffings also move the same as before (at $Ra=6.5\times10^6$), but the central plume does not sway and the flow thus becomes axisymmetric again. As shown in figure 19, the temperature profiles in the radial direction at $Ra=6.5\times10^6$ and $3.0\times10^7$ are zoomed in for a better comparison and to understand what causes the change. Although one might reckon that the meeting point of two puffs at $Ra=6.5\times10^6$ is the same as that in figure 19($e_2$), the difference can be observed clearly in zoom figures in figure 19($e_2$) and ($f_2$). The meeting point on the two sides is not the same in one period. This asymmetry of puffs finally affects the plume and makes the plume sway in different directions. However, at $Ra=3.0\times10^7$, the meeting point is the same on both sides, which means there is no symmetry breaking effects on the plume.

Further, when $Ra=5.02\times10^7$, the flow evolves into the chaotic state. According to the temperature profiles in figure 19($g_1$), there is no distinct period of the plume and the stripes become irregular. However, the puffing still appears regularly at the edge of the heated surface in figure 19($g_2$), which indicates the flow has not entered a fully-developed chaotic state. It also indicates that chaotic bifurcations firstly occur in the center plume part and then in the boundary layer at the outer regions. As shown in figure 19($g_3$), the strips appear in the form of small waves in the circumferential direction, suggesting that the puffs in the edges still have small differences in circumferential direction and it is not a symmetric state.

Additionally, two-dimensional Fourier transform (2DFT) is used to study both temporal and spatial development of the flow. While the temperature profiles in radial direction can generally describe the buoyant flow, 2DFT is applied on the temperature profiles in radial direction to obtain both frequency and wavenumber at different Rayleigh numbers. The 2DFT is performed on a long time period to ensure the accuracy of the results.

For the periodic cases, such as in figures 20(a), (b) and (c), a fundamental frequency at the peak can be found. However, the frequency value in 2DFT results is slightly different from the one obtained before. For instance, the fundamental frequency for $Ra=1.1\times10^6$ is 0.412 in figure 6(c), but is 0.415 in figure 20(a). The small difference in the fundamental frequency in the same case is mainly because the frequency in figure 6(c) is obtained by analyzing the velocity time series on a specific point ($P_1$), but the frequency obtained by 2DFT is based on the whole data set in the radial direction. For the period-doubling flow in figure 20(d), another peak at frequency $f_f/2$ can be found under a different wavenumber, which means the spatial periodicity becomes stronger for this smaller fundamental frequency. For the chaotic flow in figure 20(e), several peaks are found so that there is no fundamental frequency any more, which also indicates that more flow structures with different wavelengths appear.

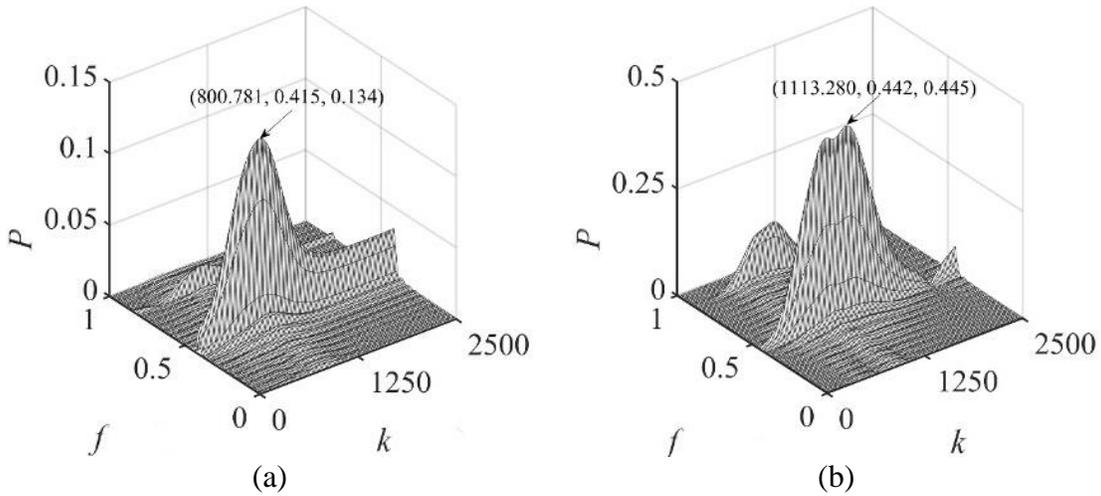

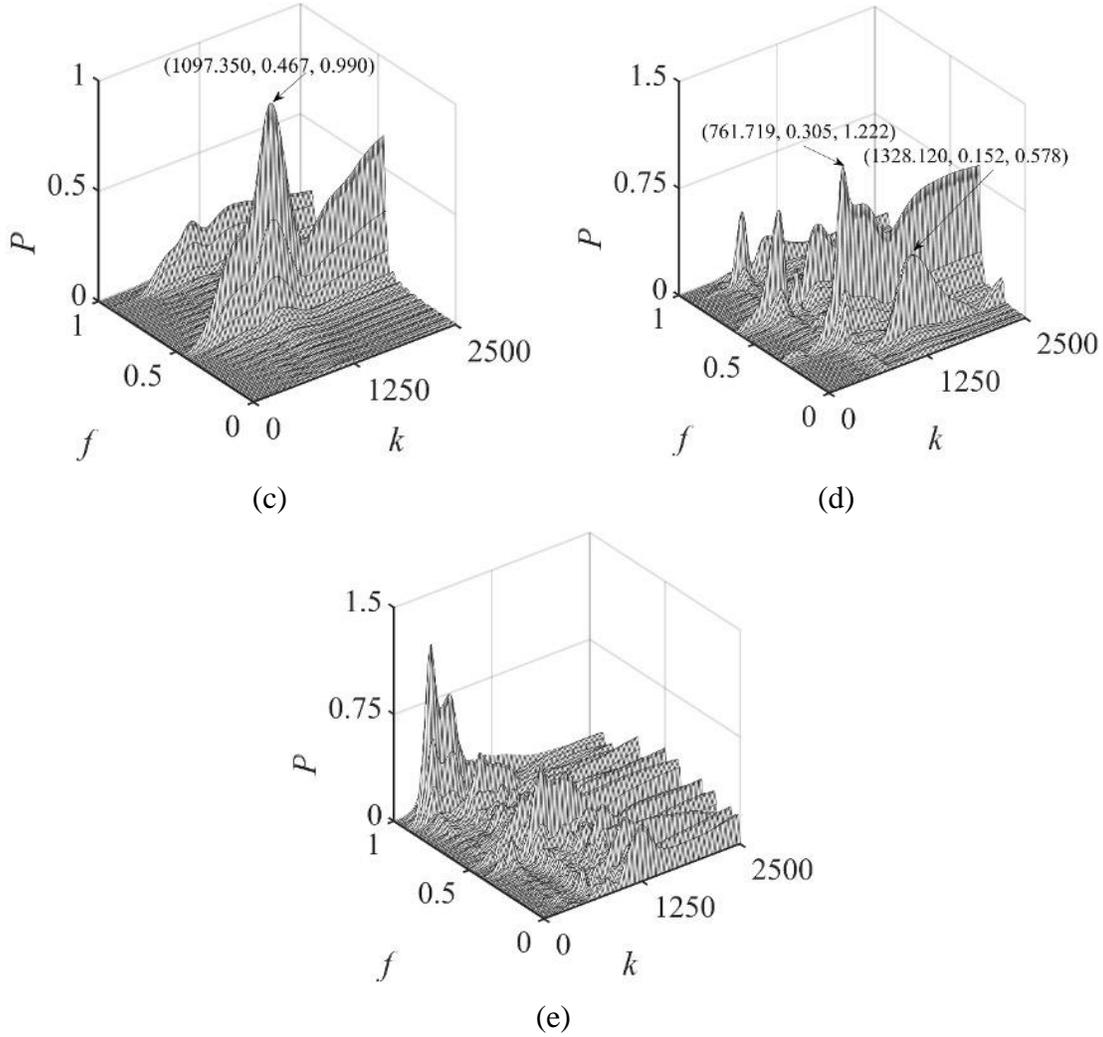

Figure 20 Two-dimensional Fourier transform of the radial temperature shown in figure 18 for (a) $Ra=1.1\times10^6$, (b) $Ra=1.9\times10^6$, (c) $Ra=2.3\times10^6$, (d) $Ra=6.5\times10^6$, and (e) $Ra=5.02\times10^7$, where $f$ and $k$ are frequency and spatial wavenumber respectively, and $P$ is spectral power.

3.2. Various states subjected to random-mode perturbations

As we discussed above, the onset of chaos usually goes through a series of bifurcations. Adding oscillations is one way to control different flow states and also control chaos (Shinbrot et al., 1993, Markman & Kedma, 1994). To examine the stability of the flow near the bifurcation point and especially the stability of different states, a direct stability analysis was performed. Random numerical perturbations were superimposed onto the boundary condition of the heated horizontal surface, which is random in both time and space. The amplitude of the random perturbations is 5% of the difference between the temperatures of the surface and ambient fluid (0.05). Furthermore, the effect of the perturbation amplitude is tested to guarantee that the response in the thermal boundary layer is in the linear regime (also see Zhao et al. 2013 for details).

Numerical results show that when $Ra=8.0\times10^5$, the flow is both stable with and without perturbations, as shown in figure 21. That is, the perturbations do not grow up and even decay when they travel downstream.

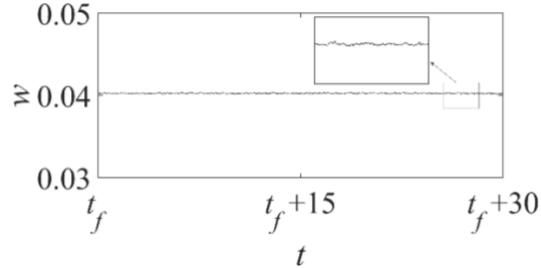

Figure 21 Time series of $z$-velocity $w$ with perturbations at $P_2$ for $Ra=8.0\times10^5$.

As shown in figure 6(a), the flow is steady for $Ra=1.0\times10^6$. Further, we also introduced random perturbations into the flow at $Ra=1.0\times10^6$. Figure 22 plots the vertical velocity time series for $Ra=1.0\times10^6$ with the random perturbations. Clearly, the flow transits into a periodic state in which the bifurcation occurs at this $Ra$. For the case without perturbations, the flow is initially periodic with the amplitude decaying with time, and it finally becomes steady after a long period of time. However, when the perturbations are added to the heating boundary, the decay of the periodic characteristics is compensated, and the flow finally evolves into a periodic state. That is, the steady flow is unstable at this point under perturbations, which drives the flow to the next transition state.

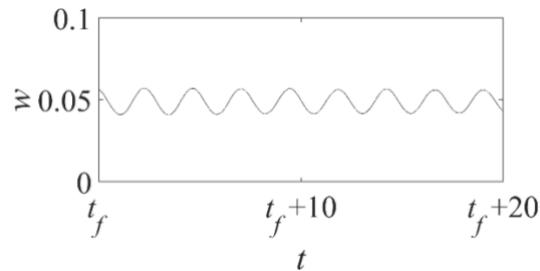

Figure 22 Time series of $z$-velocity $w$ with perturbations at $P_2$ for $Ra=1.0\times10^6$.

Further, the puffing state is also tested through introducing perturbations; that is, the calculation similar to that in figure 22 is repeated for $Ra=1.5\times10^6$. The numerical results demonstrate that the flow of the puffing state is stable.
Additionally, the flow of the rotating state at $Ra=2.0\times10^6$ is also examined. Figure 23 shows the numerical results with perturbations. Clearly, after superimposing random perturbations onto the heating boundary condition, the flow bifurcates from the rotating state in figure 23(a) to the flapping state in figure 23(b). As we discussed in section 3.1.3, the break of $O(2)$ symmetry may result in the formation of a rotating state or standing waves, depending on the initial conditions. Introducing perturbations may alter the stability of the rotating state, advancing it to other states. A series of numerical perturbation tests are therefore performed. For purpose of also understanding the influence of the perturbation amplitude, 1%, 2%, 3%, 4%, 5% and 10% of temperature difference are used as the random perturbation amplitude. It turns out that when the perturbation amplitude is large enough (>3%), the rotating state can be disturbed and it enters a flapping state in advance at the same $Ra$, implying that the rotating state is under conditional instability (Drazin 2002).

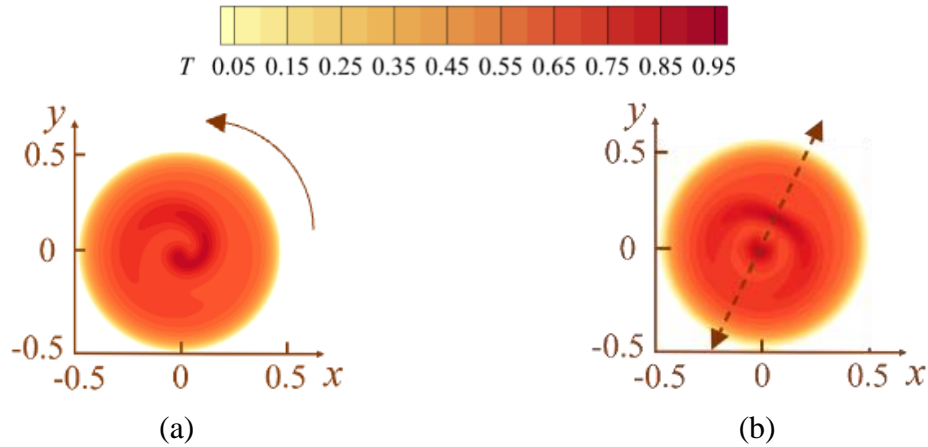

Figure 23 The $x$-$y$ plane temperature contour plot at height of $0.01D$ for $Ra=2.0\times10^6$ at equilibrium state (a) isothermal heating; (b) isothermal heating superimposed perturbations.

With further increasing $Ra$ to $6.4\times10^6$, which is near the bifurcation point from the periodic flapping state to the period-doubling flapping state, random perturbations are also added to the boundary condition. The perturbation results show that the perturbed flow remains periodic rather than entering the period-doubling state. This is probably because that the process of breaking symmetry in the spatial domain is more likely to occur than the transition in the temporal domain (from the periodic to the period-doubling), which however needs to be further investigated. When the accuracy of $Ra$ further increases to such as two or three decimal places after the decimal point, the transitional flow triggered by random perturbations may be found, which could be studied in future and is exceed the scope of the present study.

Considering the results under perturbations, we may conclude that perturbations influence the transitional flow when the flow is conditionally stable near the bifurcation point. In small $Ra$ cases, the perturbations may increase the complexity of the flow and lead the flow to bifurcate in advance to the next transition state. For example, in the rotating state, the flow is conditionally stable. When perturbations of large amplitude are introduced, the rotating state transits to the flapping state.

The whole transition route of the flow on a heated horizontal surface with the increasing of $Ra$ with and without perturbation is shown in figure 24. Without perturbation, the flow goes through a series of bifurcations and has different flow structures, from steady to periodic puffing, rotating, flapping, period doubling and finally chaos. However, with random perturbations, the flow may be affected by perturbations and bifurcates into the next transition state, such as from steady to periodic puffing and from periodic puffing to flapping without rotating state. It is worth noting that only perturbation experiments of typical states are performed because of computing time and cost.

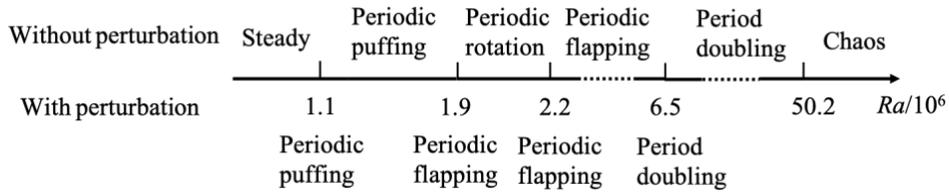

Figure 24 The transition route of the buoyant flow on a heated horizontal surface with the increasing of *Ra*, in which the dash line means the extension of the axis.

4. Conclusions

The critical transition route of the buoyant flow on an isothermally heated horizontal surface is investigated in this study using three-dimensional numerical simulations. The range of *Ra* is covered from $10^1$ to $6\times10^7$ for *Pr*=7 (water). Apart from the transition route, the stability of different states is studied using direct stability analysis, in which the influence of random perturbations on transition is examined using numerical perturbation experiments.

In the transition route of the buoyant flow, when *Ra* is less than $10^3$, the flow is under conduction dominance without presenting a distinct thermal boundary layer or starting plume, only exhibiting a heat dome structure. When *Ra* is larger, for instance in the range of $10^3<Ra<10^6$, the convective effect becomes more distinct and gradually dominates the flow, which results in the distinct rising plume while the flow is still steady and axisymmetric.

A Hopf bifurcation occurs as *Ra* is in the range of $1.0\times10^6$ and $1.1\times10^6$, resulting in a periodic puffing flow with puffings forming in the thermal boundary layer. The puffings are then entrained by the plume and convected upwards eventually. As *Ra* further increases, a symmetry-breaking Hopf bifurcation occurs for the flow between *Ra*=$1.8\times10^6$ and *Ra*=$1.9\times10^6$, where the flow enters a unique periodic rotating state and the symmetry of the flow is lost. Next, a reflection-symmetric Hopf bifurcation happens and the flow enters a periodic flapping state, which is not axisymmetric any more but is only symmetric about the vertical plane. When *Ra* is at $6.5\times10^6$, a period-doubling bifurcation occurs and the period of the flow becomes twice as large as the period of the preceding state. Finally, the flow evolves into the chaotic state when *Ra* is in the vicinity of $5.02\times10^7$.

Direct stability analysis is also conducted to understand different states. Random numerical perturbations are superimposed onto the boundary condition of the heated horizontal surface. We find that for flow regimes near bifurcation points, such as *Ra*=$1.0\times10^6$ and *Ra*=$2.0\times10^6$, the flow is conditionally stable and it bifurcates to the next stage in advance due to the perturbations. The direct stability analysis is focused on the temporal variation of the transitional flow. However, the spatial development of the perturbations, i.e., from upstream to downstream in different states, may be different. The spatial characteristics are also worth investigating in future work.

Acknowledgements
The authors would thank the National Natural Science Foundation of China for financial support (No. 11972072). The simulations were undertaken with the assistance of resources from Euler Cluster, which is supported by the Chair of Building Physics at ETH Zurich.

Supplementary materials
Supplementary movies are available.

Declaration of interests
The authors claim no conflict of interest.

Data availability statement
All data are involved in the article.